\documentclass[manuscript,screen,nonacm, 11pt]{acmart}

\newcommand\vldbavailabilityurl{https://github.com/tomereven/Prefix-Filter}

\makeatletter
\renewcommand{\paragraph}{%
  \@startsection{paragraph}{4}{\z@}%
  {-.5\baselineskip \@plus -2\p@ \@minus -.2\p@}{-3.5\p@}%
  {\normalfont\normalsize\bfseries}%
}

\renewcommand\subsubsection{\@startsection{subsubsection}{3}{\z@}%
   {-.75\baselineskip \@plus -2\p@ \@minus -.2\p@}%
   {.25\baselineskip}%
   {\bfseries\section@raggedright}}
\makeatother

\input{include.tex}

\usepackage{amsmath} \usepackage{amsthm}
\newtheorem{theorem}{Theorem}%

\newtheorem{corollary}[theorem]{Corollary}
\newtheorem{conclusion}[theorem]{Conclusion}
\newtheorem{claim}[theorem]{Claim}
\newtheorem{remark}[theorem]{Remark}
\newtheorem{proposition}[theorem]{Proposition}

\newtheorem{definition}[theorem]{Definition}
\newtheorem{invariant}[theorem]{Invariant}

\newtheorem{observation}[theorem]{Observation}

\newcommand{\etal}{\textit{et al.}\xspace}

\newcommand{\filter}[1]{{#1}\xspace}

\newcommand{\PF}{\filter{prefix filter}}
\newcommand{\VQF}{\filter{vector quotient filter}}

\newcommand{\CF}{\filter{cuckoo filter}}

\newcommand{\BF}{\filter{Bloom filter}}
\newcommand{\BBF}{\filter{blocked Bloom filter}}

\newcommand{\BBFx}{\filter{BBF-Flex}}
\newcommand{\BE}{\filter{BE filter}}

\newcommand{\spare}{\filter{spare}}

\newcommand{\fp}{\textnormal{\textsf{fp}}}

\newcommand{\yes}{\textnormal{``\textsf{Yes}''}\xspace}
\newcommand{\no}{\textnormal{``\textsf{No}''}\xspace}

\newcommand{\UU}{\mathcal{U}}

\newcommand{\ZZ}{\mathcal{Z}}

\newcommand{\DD}{\mathcal{D}}

\newcommand{\su}{s}

\newcommand{\Bin}[2]{\mathcal{B}_{#1}^{#2}}

\newcommand{\FP}{\text{FP}}

\renewcommand{\epsilon}{\varepsilon}
\newcommand{\eps}{\varepsilon}

\DeclarePairedDelimiter{\floor}{\lfloor}{\rfloor}

\newcommand{\size}[1]{\ensuremath{\left|#1\right|}}
\newcommand{\set}[1]{\left\{ #1 \right\}}
\newcommand{\lset}[1]{\left\langle #1 \right\rangle}

\newcommand{\brak}[1]{\left(#1\right)}
\newcommand{\expectation}[2]{\mathbb{E}_{#1}\left[ #2 \right]}
\newcommand{\Exp}[1]{\mathbb{E}\left[ #1 \right]}
\newcommand{\variance}[1]{\mathbb{V}\left[ #1 \right]}
\renewcommand{\Pr}[1]{{\mathrm{Pr}}\left[ #1 \right]}

\DeclareMathOperator{\query}{\textnormal{\textsf{query}}}

\DeclareMathOperator{\bin}{\textnormal{\textsf{bin}}}

\DeclareMathOperator{\ins}{\textnormal{\textsf{insert}}}

\DeclareMathOperator{\hheader}{\textnormal{header}}
\DeclareMathOperator{\bbody}{\textnormal{body}}
\DeclareMathOperator{\PD}{\textnormal{\textsf{PD}}}

\DeclareMathOperator{\Select}{\textnormal{\textsf{Select}}}
\DeclareMathOperator{\Rank}{\textnormal{\textsf{Rank}}}
\DeclareMathOperator{\rank}{\mathsf{rank}}

\newcommand{\cProbVar}{\frac{200\pi k}{0.99\cdot n}} %
\newcommand{\cgam}{\frac{1}{\sqrt{2\pi k}}}
\newcommand{\cdom}{\sqrt{2\pi k}}

\newif\ifnotes
\notestrue
\ifnotes
	\newcommand{\tomer}[1]{{\ifnotes \scriptsize \textcolor{red}{Tomer: {#1}} \fi}}
	\newcommand{\adam}[1]{{\ifnotes \scriptsize \textcolor{green}{Adam: {#1}} \fi}}
	\newcommand{\guy}[1]{\ifnotes {\noindent \scriptsize  \textcolor{blue} {Guy: {#1}}} \fi{}}
\else
	\newcommand{\tomer}[1]{}
	\newcommand{\guy}[1]{}
	\newcommand{\adam}[1]{}
\fi

\begin{document}
\title{Prefix Filter: Practically and Theoretically Better Than Bloom}

\author{Tomer Even}
\affiliation{%
    \institution{Tel Aviv University}
    \country{Israel}
}
\author{Guy Even}
\affiliation{%
    \institution{Tel Aviv University}
    \country{Israel}
}
\author{Adam Morrison}
\affiliation{%
    \institution{Tel Aviv University}
    \country{Israel}
}
\begin{abstract}

    Many applications of approximate membership query data structures, or \emph{filters}, require only an \emph{incremental} filter that supports insertions but not deletions.
    However, the design space of incremental filters is missing a ``sweet spot'' filter that combines space efficiency, fast queries, and fast insertions.
    Incremental filters, such as the Bloom and blocked Bloom filter, are not space efficient.
    Dynamic filters (i.e., supporting deletions), such as the cuckoo or vector quotient filter, are space efficient but do not exhibit consistently fast insertions and queries.

    In this paper, we propose the \emph{\PF}, an incremental filter that addresses the above challenge: (1) its space (in bits) is similar to state-of-the-art dynamic filters; (2) query throughput is high and is comparable to that of the \CF; and (3) insert throughput is high with overall build times faster than those of the \VQF and \CF by $1.39\times$--$1.46\times$ and $3.2\times$--$3.5\times$, respectively. %
    We present a rigorous analysis of the \PF that holds also for practical set sizes (i.e., $n=2^{25}$). The analysis deals with the probability of failure, false positive rate, and probability that an operation requires accessing more than a single cache line.

\end{abstract}

\maketitle

\ifdefempty{\vldbavailabilityurl}{}{
    \vspace{.3cm}
    \begingroup\small\noindent\raggedright\textbf{Artifact Availability:}\\
    The source code, data, and/or other artifacts have been made available at \url{\vldbavailabilityurl}.
    \endgroup
}
\section{Introduction}

\paragraph{What is a filter?}
An approximate membership query (AMQ) data structure, or \emph{filter}~\cite{bloom1970spacetime}, is a data structure for approximately maintaining a set of keys.
A filter is specified by three parameters: $\UU$ - the universe from which keys are taken, $n$ - an upper bound on the number of keys in the set, and $\eps$ - an upper bound on the false positive rate.
A query for a key in the set must output \yes, while the output of a query for a key not in the set is \yes with probability at most $\eps$.

\paragraph{What are filters used for?}
The main advantage of filters over data structures that offer exact representations of sets (i.e., dictionaries/hash tables) is the reduced space consumption which leads to increased throughput.
Exact representation of a set of $n$ keys from a universe of size $u$ requires at least $\log_2(u/n)$ bits per key.%
\footnote{Cuckoo hashing, for example, requires in practice $2\log u$ bits per keys~\cite{dblp:journals/jal/paghr04,10.1145/2592798.2592820} (throughput deteriorates when a cuckoo hash table is more than half full).
}
On the other hand, a filter requires at least $\log_2(1/\eps)$ bits per key~\cite{carter1978exact}.
Namely, the space per key in a filter need not depend on the universe size.
The reduced space allows for storing the filter in RAM, which leads to higher throughput.

The typical setting in which filters help in increasing performance is by saving futile searches in some slow (compared to a filter) data store. To reduce the frequency of \emph{negative queries} (i.e., queries for keys that are not in the data store), every lookup is preceded by a query to a filter.
A negative filter response guarantees that the key is not in the data store, and saves the need for a slow futile access to the data store.
The filter reduces the frequency of negative queries by a
factor of at least $1/\eps$, where $\eps$ is the upper bound on
filter's false positive rate.
It is therefore important to
design filters with high throughput specifically for negative queries.

Examples of applications that employ the above paradigm include databases~\cite{HashingMethods}, storage systems~\cite{LSM-Survey}, massive data processing~\cite{Synopses-for-Massive-Data}, and similar domains.
Filter efficiency can dictate overall performance, as filters can consume much of the system's execution time~\cite{SplinterDB,Chucky,SlimDB} and/or memory~\cite{dayan2017monkey,PebblesDB,PartIndexFilter}.

\paragraph{Importance of incremental filters.}
Many filter use cases require only an \emph{incremental} filter, which supports only insertions, and do not require a \emph{dynamic} filter that also supports deletions.
For instance, numerous systems~\cite{KVstoreAnalysis,RocksDBAnalysis,conway_et_al:LIPIcs:2018:9043,SplinterDB,dayan2017monkey,PebblesDB,rocksdb,LSM-SSD,Chucky,Dostoevsky,chang2008bigtable,Cassandra,bLSM,cLSM} use log-structured merge (LSM) tree~\cite{DBLP:journals/acta/ONeilCGO96} variants as their storage layer. In these systems, the LSM tree data in secondary storage consists of multiple immutable files called \emph{runs}. Each run has a corresponding filter in main memory, which is built when the run is created and is subsequently only queried (as runs are immutable).
Consequently, these systems only need an incremental filter.%

\paragraph{The problem.}
The design space of incremental filters is missing a ``sweet spot'' filter that combines space efficiency, fast queries, and fast insertions. Incremental filters, such as the \BBF~\cite{putze2010cache}, are fast but not space efficient. For a typical error rate of $\eps=2.5$\%, the \BBF uses up to $1.6\times$ the space of the information theoretic lower bound of $n \log_2(1/\eps)$ bits~\cite{carter1978exact}.
State-of-the-art dynamic filters (which can serve as incremental filters)
are space efficient but do not exhibit consistently fast insertions and queries.
The \CF~\cite{dblp:conf/conext/fanakm14} has the fastest queries, but its insertions slow down to a crawl as filter occupancy grows close to $n$---e.g., we observe a $27\times$ slowdown (see~\cref{sec:evaluation}).
In the \VQF~\cite{pandey2021vector}, insertion throughput does not decline as dramatically as filter occupancy grows, but query throughput is lower than the \CF.
When filter occupancy is high, the \CF's query throughput is about
$52\%/79\%$ faster for negative/positive queries, respectively.
Finally, %
in the quotient filter~\cite{bender2012thrash} insertions and queries are slower than in the \VQF.

\paragraph{The prefix filter.}
In this paper, we propose the \PF, an incremental filter that addresses the above challenge:
\begin{enumerate*}[label=(\arabic*)]
    \item
    its space is close to optimal, similarly to state-of-the-art dynamic filters;
    \item it has fast queries, comparable to those of the \CF; and
    \item it has fast insertions, with overall build time faster than those of the \VQF and \CF by $1.39\times$--$1.46\times$ and $3.2\times$--$3.5\times$, respectively.
\end{enumerate*}
Compared to the \BBF (BBF), the \PF is slower, but far more space efficient at low error rates.
The \PF exhibits a good trade-off in many cases, where the BBF's high space usage is limiting but dynamic filter speeds (of $<35$\,nanoseconds/query~\cite{pandey2021vector}) are not.

The \PF shares the common high-level structure of modern dynamic filters (hence, its space efficiency) but it exploits the lack of deletion support to simultaneously obtain fast queries and insertions.
This common structure is a hash table that stores short hashes of the keys, called \emph{fingerprints}, with filters differing in how they resolve hash table collisions. For example, the \CF employs cuckoo hashing~\cite{dblp:journals/jal/paghr04}, whose insertion time depends on table occupancy, whereas the \VQF uses power-of-two-choices hashing, whose insertions are constant-time.
Crucially, however, these collision resolution schemes result in every filter query performing \emph{two} memory accesses
for negative queries, typically incurring two cache misses (because each memory access is to a random address).

The \PF uses a novel solution to the collision resolution problem, in which typically only a \emph{single} cache line is accessed per filter operation.
Our starting point is the theoretical work of~\cite{bercea2019fullydynamic,DBLP:conf/swat/BerceaE20}, which describes a dynamic space-efficient filter that we call the \BE. In a nutshell,
the \BE is a two-level hash table for fingerprints, where the first level stores most of the fingerprints.
In contrast to cuckoo or power-of-two-choice hashing, insertions try to insert the new key's fingerprint into a \emph{single} bin in the first level. If that bin is full, the fingerprint is inserted into the second level, called the \emph{spare}.
Bins are implemented with a space-efficient data structure called a \emph{pocket dictionary} ($\PD$), which employs the Fano-Elias encoding~\cite{fano1971number,elias1974efficient,carter1978exact}.
The \spare, however, can be any hash table scheme that reaches high occupancy (about $95\%$) with low failure probability.

The \BE always accesses two cache lines per negative query because the key is searched for in both levels.
The \PF solves this problem by \emph{choosing} which key to forward to the \spare (upon insertion to a full bin) instead of simply forwarding the currently inserted key.
Specifically, upon insertion to a full bin, the \PF forwards the maximal fingerprint among the fingerprints stored in the full bin plus the new key.
This maintains the invariant that each bin stores a maximal prefix of the fingerprints that are mapped to it.
We refer to this invariant as the \emph{Prefix Invariant}.
The Prefix Invariant saves the need of forwarding a negative query to the \spare if the searched fingerprint should have resided in the stored prefix.
This simple idea results in most queries completing with \emph{one} cache line access.
We prove that the probability that a query searches the \spare is at most $\frac1{\sqrt{2\pi k}}$, where $k$ is the capacity of the first-level bins~(\cref{sec:prefix-ma}). In our prototype with $k=25$, roughly $92\%$ of negative queries access one cache line.

To further improve the filter's speed, we design a novel implementation of the pocket dictionary structure that implements a first-level bin.
Our implementation exploits vector (SIMD) instructions to decode the Elias-Fano encoding~\cite{elias1974efficient,fano1971number,carter1978exact} used in the PD without performing expensive $\Select$~\cite{pibiri2021rank,DBLP:journals/corr/MulaKL16,DBLP:journals/corr/PandeyBJ17} computations.
Our implementation may be of independent interest, as the PD (or variants) is at the heart of other filters~\cite{einziger2017tinyset, pandey2021vector}.

Another challenge in achieving a practical filter based on the \BE has to do with the size and implementation of the \spare.
The asymptotic analysis of the \BE (and~\cite{arbitman2010backyard}) proves that the number of keys stored in the \spare is negligible, and it is therefore suggested to implement the \spare using a dictionary.
In practice, however, $6\%$--$8\%$ of the keys are forwarded to the \spare because the asymptotics ``start working'' for impractically large values of $n$.
Therefore, a practical implementation needs to provide a high throughput implementation of the \spare that is also space efficient.
To this end, the \PF implements the \spare using a filter.

We rigorously analyze the \PF. The mathematical analysis of the false positive rate, size of the spare, and the fraction of queries that access one cache line does not ``hide'' any constants. Therefore, this analysis holds both for practical parameter values (e.g., $n$ as small as $2^{25}$) and asymptotically.
This is in contrast to filters such as the \CF, whose theoretical asymptotic space requirements are worse than the \BF~\cite{dblp:conf/conext/fanakm14}.
We also empirically demonstrate the \PF's high throughput and space efficiency.

\paragraph{Contributions.} The contributions of this paper are summarized below.
\begin{itemize}[leftmargin=*]
    \item \textbf{The \PF:}
          The \PF is space-efficient, supports insertions and membership queries, and supports sets of arbitrary size (i.e., not restricted to powers of two).
    \item \textbf{Rigorous analysis:} The \PF is accompanied by a mathematical analysis that proves upper bounds on the
          probability of failure, false positive rate, and probability that an operation requires accessing more than a single cache line.
    \item \textbf{Implementation:}
          We implement the \PF in C++. The implementation includes an efficient PD that avoids expensive computations such as $\Select$ computations (\cref{sec:optimize}). The \PF code is available at \url{https://github.com/tomereven/Prefix-Filter}.
    \item\textbf{Evaluation:}
          We show that the \PF's build time (from empty to full) is faster than that of the \VQF and \CF by $1.39\times$--$1.46\times$ and $3.2\times$--$3.5\times$, respectively.
          Throughput of negative queries in the \PF is comparable to that of the \CF and faster than the \VQF by about $1.38\times$--$1.46\times$ (\cref{sec:evaluation}).
\end{itemize}

\section{Preliminaries}\label{sec:back}

\paragraph{Problem statement: Filters.}
A \emph{filter} is a data structure used for approximate representation of a set a of \emph{keys}.
A filter receives three parameters upon creation: (1) $\UU$, the universe from which keys are taken; (2) $n$, an upper bound on the cardinality of the set; and (3) $\eps$, an upper bound on the false positive rate.
An \emph{incremental} filter supports two types of operations: $\ins(x)$ and $\query(x)$.
An $\ins(x)$ adds $x$ to the set; we assume $x$ is not already in the set. We denote by $\DD$ the set defined by the sequence of distinct $\ins$ operations.
Responses for queries allow one-sided errors: if $x \in \DD$, then $\query(x)=\text{\yes}$; if $x\not\in \DD$, then $\Pr{\query(x)=\text{\yes}}\leq \eps$.
The probability space is over the randomness of the filter (e.g., choice
of hash function) and does not depend on the set or the queried
key.

\paragraph{Types of queries.}
A query for a key in the set is a \emph{positive query}.
A query for a key not in the set is a \emph{negative query}.
\subsection{Terminology}

In this section we define various terms used throughout the paper.
The reader may wish to skip this section and return to it upon encountering unfamiliar terms and notation.

\paragraph{Notation.}
All logarithms in this paper are base 2.
For $x>0$, we use $[x]$ to denote the set $\set{0,1,\ldots,\floor{x}-1}$.
We use $\exp(x)$ to denote $e^x$.

\paragraph{Dictionary.} A \emph{membership-dictionary} (or \emph{dictionary}) is a data structure used for exact representation of sets $\DD\subseteq\UU$.
Namely, responses to membership queries are error-free.

\paragraph{Bins.} Dictionaries and filters often employ a partitioning technique that randomly maps keys to \emph{bins}.
This partitioning reduces the size of the problem from the size of the set to (roughly) the average size of a bin.
Each bin is a membership-dictionary that stores keys mapped to the same bin.
The \emph{capacity} of a bin is the maximum number of keys it can store. A bin is \emph{full} if it contains the maximal number of keys it can store.
A bin \emph{overflows} if one attempts to insert a new key to it while it is full.

\paragraph{Load vs. load factor}
The \emph{load} of a filter is the ratio between the cardinality of the set stored in the filter and its maximum size, i.e., $|\DD|/n$.
The \emph{load factor} of a table of $m$ bins, each of capacity $k$, is the ratio between the total number of keys stored in the bins and $m \cdot k$.
We purposefully distinguish between the concepts of load and load factor. ``Load'' is well-defined for every filter, whereas ``load factor'' is defined only for a table of bins---and a filter is not necessarily such a table, and even when it is, typically $mk > n$ (see~\cref{sec:related}).

\paragraph{Failure.} A filter \emph{fails} if it is unable to complete an $\ins(x)$ operation although $|\DD| < n$.
Filter designs aspire to minimize failure probability.

\paragraph{Fingerprint.}
The fingerprint $\fp(k)$ of a key $k$ is the image of a random hash function.
The length of a fingerprint is usually much shorter than the length of the key.

\paragraph{Quotienting.}
Quotienting is a technique for reducing the number of bits needed to
represent a key~\cite{Knuth}.
Let $Q,R>0$.
Consider a universe $[Q]\times [R]$.
Consider a key $x=(q,r)\in[Q]\times [R]$. We refer to $r$ as the \emph{remainder} of $x$ and refer to $q$ as the \emph{quotient} of $x$.
In quotienting, a set $\DD\subset [Q]\times [R]$ is	stored in an array $A[0:(Q-1)]$ of ``lists'' as follows:
a key $(q,r)\in[Q]\times [R]$ is inserted by adding $r$ to the list $A[q]$.

\section{Related Work} \label{sec:related}

There are two main families of filter designs: \emph{bit-vector} and \emph{hash table of fingerprints} designs.
We compare filter designs that have been implemented according to their space requirements (bits per key), number of cache misses incurred by a negative query, and (for hash tables of fingerprints) the maximal load factor of the underlying hash table, beyond which the filter might occasionally fail.
\Cref{table:the-space} summarizes the following discussion and the properties of the \PF's.

\begin{table}
    \newlength{\myl}
    \settowidth{\myl}{$\pmb{\frac{(1 + \gamma)}{\alpha}\cdot(\log(1/\eps) +2) + \frac{\gamma}{\alpha}}$ }

    \begin{tabular}{l|p{\the\myl}lc}
        \toprule
        \textbf{Filter}       & \textbf{Bits Per Key}                                                             & \textbf{CM$/$NQ}       & \thead{\textbf{Max. Load} \\  \textbf{Factor} } \\
        \midrule
        \textsf{BF}           & $1.44\cdot \log(1/\eps)$                                                          & $\approx 2$            & -                         \\
        \textsf{BBF}          & $\approx10$--$40$\% above \textsf{BF}                                             & 1                      & -                         \\
        \midrule
        \textsf{CF}$^\dagger$ & $1/\alpha \, (\log (1/\eps) +3)$                                                  & 2                      & $94\%$                    \\
        \textsf{VQF}          & $1/\alpha \, (\log (1/\eps)+2.9)$                                                 & 2                      & $94.5\%$                  \\
        \midrule
        \textbf{\textsf{PF}}  & $\pmb{\frac{(1 + \gamma)}{\alpha}\cdot(\log(1/\eps) +2) + \frac{\gamma}{\alpha}}$ & $\pmb{\leq 1+2\gamma}$ & $\pmb{100\%}$             \\
        \bottomrule
    \end{tabular}

    \caption{Comparison of practical filters' space requirements, average cache misses per negative query (CM/NQ), and maximal load factor (for hash tables of fingerprints).
        For the \PF, $\gamma \triangleq \frac{1}{\sqrt{2\pi k}}$, where $k$ denotes the capacity of its hash table bins. The space formula is derived in~\cref{sec:the-pf:params}.
        \newline $^\dagger$ We assume (throughout the paper) a cuckoo filter with bins of 4 fingerprints and 3 bits of space overhead (which is faster than other \CF variants~\cite{dblp:conf/conext/fanakm14,pandey2021vector}) and $n < 2^{64}$, as asymptotically CF fingerprints are not constant in size~\cite{dblp:conf/conext/fanakm14}.
    }
    \label{table:the-space}
\end{table}

In a \emph{bit-vector} design, every key is mapped to a subset of locations in a bit-vector, and is considered in the set if all the bits of the bit-vector in these locations are set. %
Bit-vector filters, such as the Bloom~\cite{bloom1970spacetime} and blocked Bloom~\cite{putze2010cache} filter (BF and BBF), have a non-negligible multiplicative space overhead relative to the information theoretic minimum.
For example, a \BF uses $1.44\times$ bits per key than the minimum of $\log(1/\eps)$.

The \emph{hash table of fingerprints} design paradigm was proposed by
Carter~\etal~\cite{carter1978exact}.
In this paradigm, keys are hashed to short fingerprints which are inserted in a hash table. Designs differ in (1) table load balancing technique (e.g., cuckoo~\cite{dblp:conf/conext/fanakm14,breslow2018morton}, Robin Hood~\cite{bender2012thrash}, or power-of-two-choices~\cite{pandey2021vector} hashing); and (2) fingerprint storage techniques (e.g., explicitly storing full fingerprints~\cite{dblp:conf/conext/fanakm14}, using quotienting to explicitly store only fingerprint suffixes~\cite{bender2012thrash}, lookup tables~\cite{arbitman2010backyard}, or using the Fano-Elias encoding~\cite{pandey2021vector,einziger2017tinyset}).

Modern hash table of fingerprint designs use essentially the same space.
The hash table explicitly stores $\log(1/\eps)$ fingerprint bits, resulting in space requirement of $(\log(1/\eps + c)) \cdot 1/\alpha$ bits per key, where $c$ depends on the (per-key) overhead of hash table and encoding metadata, and $\alpha$ is the hash table's load factor.
Ideally, $\alpha$ can reach $1$, but load balancing techniques often have a maximal feasible load factor, $\alpha_{\max}$, beyond which insertions are likely to fail (or drastically slow down)~\cite{dblp:conf/conext/fanakm14,breslow2018morton,pandey2021vector,bender2012thrash}.
The relevant filters thus size their tables with $n/\alpha_{\max}$ entries, so that at full filter load, the load factor is $\alpha_{\max}$.

Viewed as incremental filters, existing hash table of fingerprints filters have limitations that are solved by the \PF (PF).
In the Cuckoo filter (CF)~\cite{dblp:conf/conext/fanakm14}, insertions slow down dramatically (over $27\times$) as load increases, resulting in slow build times.
The \VQF (VQF) has stable insertion speed, but in return, its queries are slower than the \CF's~\cite{pandey2021vector}.
Negative queries in the cuckoo and vector quotient filters always access two table bins and thus incur two cache misses.
The Morton~\cite{breslow2018morton} and quotient~\cite{bender2012thrash} filters both have insertions and queries that are slower than those of the \VQF.

The \PF has both fast insertions and queries that typically incur only one cache miss, and comparable space requirements to other hash table of fingerprints designs.
Like the \VQF and TinySet~\cite{einziger2017tinyset}, the \PF's hash table bins are succinct data structure based on the Fano-Elias encoding~\cite{fano1971number,elias1974efficient}.
While TinySet queries also access a single bin, its elaborate encoding has no practical, efficient implementation.
In contrast, we describe a practical bin implementation based on vector (SIMD) instructions.

\section{The Prefix Filter} \label{sec:the-pf}

The \PF is a ``hash table of fingerprints'' incremental filter design, distinguished by the property that its load balancing scheme requires queries and insertions to typically access only a \emph{single} hash table bin, even at high load factors.
This property translates into filter operations requiring a single cache miss (memory access) in practical settings, where the parameters are such that each \PF bin fits in one cache line.

\paragraph{High-level description.}
The \PF is a two-level structure, with each level storing key fingerprints. The first level, called the \emph{bin table}, is an array of bins to which inserted fingerprints are mapped~(\cref{sec:the-pf:bins}). Each bin is a dictionary with constant-time operations, whose capacity and element size depend on the desired false positive rate.
The second level, called the \emph{\spare}, is an incremental filter whose universe consists of key fingerprints~(\cref{sec:the-pf:spare}).

The bin table stores most fingerprints.
The \spare stores (i.e., approximates) the multiset
of fingerprints which do not ``fit'' in the bin table; specifically, these are fingerprints whose bins are full and are larger than all fingerprints in their bin.
We prove that typically, at most $\frac{1.1}{\sqrt{2\pi k}}$ of the fingerprints are thus \emph{forwarded} for storage in the \spare,
where $k$ is the capacity of the bins~(\cref{sec:spare-analysis}).

The crux of the algorithm is its above policy for choosing which fingerprints to store in the \spare.
This policy maintains the invariant that each bin stores a maximal prefix of the fingerprints that are mapped to it, which allows queries to deduce if they need to search the \spare. We prove that as a result, queries need to search the \spare only with probability at most $\frac1{\sqrt{2\pi k}}$ (\cref{sec:prefix-ma}), so most queries complete with a single bin query.

Due to the influence of $k$ on the \spare's dataset size and its probability of being accessed, the \PF needs $k$ to be as large as possible while simultaneously fitting a bin within a single cache line. To this end, our \PF prototype implements bins with a succinct dictionary data structure called a \emph{pocket dictionary}~(\cref{sec:optimize}).
Conceptually, however, the \PF can work with any dictionary, and so in this section, we view a bin simply as an abstract dictionary datatype.

\subsection{First Level: Bin Table} \label{sec:the-pf:bins}

The bin table consists of an array of $m$ bins.
Each bin has capacity $k$ and holds elements from $[s]$.
The values of $m$, $k$, and $s$ are determined by the dataset size $n$ and desired false positive rate $\eps$, specified at filter creation time.
We defer discussion of these value settings to~\cref{sec:the-pf:params}.

Keys are mapped to \emph{fingerprints} with a universal hash function $\FP$.
The bin table applies quotienting to store these fingerprints without needing to explicitly store all bits of each fingerprint.
Specifically, we view the fingerprint of a key $x$ as a pair $\FP(x)=(\bin(x), \fp(x))$, where $\bin(x) : \UU \rightarrow [m]$ maps $x$ to one of the $m$ bins, which we call \emph{$x$'s bin}, and $\fp(x) : \UU \rightarrow [s]$ maps $x$ to an element in the bin's element domain, which we call $x$'s \emph{mini-fingerprint}.
At a high level, the bin table attempts to store $\fp(x)$ in $\bin(x)$ (abusing notation to identify a bin by its index).

\paragraph{Insertion.}
\Cref{alg:ins pf} shows the \PF's insertion procedure. An $\ins(x)$ operation first attempts to insert $\fp(x)$ into $\bin(x)$. If $\bin(x)$ is full, the operation applies an eviction policy that forwards one of the fingerprints mapping to $\bin(x)$ (possibly $\fp(x)$ itself) to the \spare. The policy is to forward the maximum fingerprint among $\FP(x)$ and the fingerprints currently stored in $\bin(x)$.
The full maximum fingerprint can be computed from the mini-fingerprints in the bin because they all have the same bin index.
If $\FP(x)$ is not forwarded to the \spare, then $\fp(x)$ in inserted into $\bin(x)$ in place of the evicted fingerprint.
Finally, the bin is marked as \emph{overflowed}, to indicate that a fingerprint that maps to it was forwarded to the spare.

    {%
        \begin{algorithm}[t]
            \caption{Prefix filter insertion procedure}\label{alg:ins pf}
            \KwIn{$x \in \UU$.}
            \medskip
            \If {$\bin(x)$ is not full}{
                $\bin(x).\ins(\fp(x))$ \;
            } \Else {
                $\fp_{\max} \gets \max \left\{ \fp \, | \, \fp \in \bin(x) \right\}$ ; \Comment{\small Computed in $O(1)$ time (\cref{sec:pd-exts})}
                \If{$\fp(x) > \fp_{\max}$}{
                    \spare.$\ins(\FP(x))$ \;
                } \Else {
                    \spare.$\ins((\bin(x), \fp_{\max}))$ \;
                    replace $\fp_{\max}$ with $\fp(x)$ in $\bin(x)$ \;
                }
                $\bin(x).\mathsf{overflowed} = \mathsf{TRUE}$ \;
            }
        \end{algorithm}

        The \PF's eviction policy is crucial for the filter's operation. It is what enables most negative queries to complete without searching the \spare, incurring a single cache miss. The eviction policy maintains the following \emph{Prefix Invariant} at every bin:
        \begin{invariant}[Prefix Invariant]\label{inv:prefix}
            For every $i \in [m]$, let $$\DD_i \triangleq \{\fp(x) \, | \, x \in\DD\text{ and } \bin(x) = i\}$$
            be the mini-fingerprints of dataset keys whose bin index is $i$ inserted into the filter so far (i.e., including those forwarded to the \spare). Let $Sort(\DD_i)$ be the sequence obtained by sorting $\DD_i$ in increasing lexicographic order. Then bin $i$ contains a prefix of $Sort(\DD_i)$.
        \end{invariant}

        Maintaining the~\nameref{inv:prefix} makes it impossible for the \PF to support deletions efficiently. The problem is that when the maximum mini-fingerprint is deleted from an overflowed bin, there is no way to extract the fingerprint that should ``take its place'' from the \spare.

        \paragraph{Query.}
        The~\nameref{inv:prefix} enables a $\query(x)$ operation that does not find $x$ in $\bin(x)$ to deduce whether it needs to search the \spare for $\FP(x)$. This search need to happen only if $\bin(x)$ has overflowed and $\fp(x)$ is greater than all mini-fingerprints currently stored in the bin.
        \Cref{alg:q pf} shows the pseudo code. Our bin implementation supports finding the bin's maximum mini-fingerprint in constant time~(\cref{sec:pd-exts}).

        \begin{algorithm}[t]
            \caption{Prefix filter query procedure}\label{alg:q pf}
            \setcounter{AlgoLine}{0}
            \KwIn{$x \in \UU$.}
            \KwOut{Output \yes if $\fp(x)$ was previously inserted into $\bin(x)$. Otherwise, output \no.}
            \medskip
            \If {$\bin(x).\mathsf{overflowed}$ \textbf{ and } $(\fp(x)> \max \left\{ \fp \, | \, \fp \in \bin(x) \right\}$}{
                \Return \ \spare.$\query(\FP(x))$ \;
            }
            \Else{
                \Return $\bin(x).\query(\fp(x))$ \;
            }
        \end{algorithm}
    }

\subsection{Second Level: The Spare} \label{sec:the-pf:spare}

The \spare can be any incremental filter for the universe $\UU'$ of key fingerprints (i.e., $\UU' = \left\{ FP(x) \, | \, x \in \UU \right \}$).
\Cref{sec:the-pf:spare-params} describes how the \spare is parameterized, i.e., which dataset size and false positive rate it is created with. Determining the \spare's dataset size, in particular, is not trivial: the number of fingerprints forwarded to the spare over the lifetime of the \PF is a random variable, but the \PF must provide some bound to the \spare upon creation.
\Cref{sec:the-pf:which-spare} discusses how the \spare's speed and space requirement affect the \PF's overall speed and space requirement.

\subsubsection{Spare Parameters} \label{sec:the-pf:spare-params}

Here we describe how the \PF chooses the \spare's dataset size bound and false positive rate, both of which must be chosen when the filter is constructed.

\paragraph{Dataset size.}
The problem with specifying the \spare's dataset size, denoted $n'$, is that the number of fingerprints forwarded to the \spare is a random variable.
If we overestimate $n'$, the \spare will consume space needlessly.
But if we underestimate $n'$, \spare insertions might fail if it receives more than $n'$ insertions---causing the \PF to fail.
(This is akin to the situation in other hash table-based filters, which might fail with some probability.)

To set $n'$ to a value that yields a low \PF failure probability, we (1) prove that the expectation of the random variable $X$ whose value is the number of fingerprints forwarded to the \spare is
$\mathbb{E}[X] \approx n/\sqrt{2\pi k}$ and (2) bound the probability that $X$ deviates from its expectation by a factor of $1+\delta$~(\cref{sec:spare-analysis}).
Based on this bound, we suggest setting $n' = 1.1 \cdot \mathbb{E}[X]$,
which yields a \PF failure probability (due to $X > n'$) of at most $\cProbVar$.

\paragraph{Spare false positive rate.}
The \spare's false positive rate $\eps'$ only marginally affects the \PF's false positive rate.
We prove in~\cref{sec:filter-fpp} that the prefix filter's false positive rate is bounded by $\frac{\alpha\cdot k}{s} + \frac{1}{\sqrt{2\pi k}} \eps'$.
Therefore, the main determining factor for the \PF's false positive rate is the choice of bin table parameters (see further discussion in~\cref{sec:the-pf:params}).

\subsubsection{Impact of Spare Speed and Space} \label{sec:the-pf:which-spare}

Speed-wise, queries access the \spare infrequently.
Less than a $1/\sqrt{2 \pi k}$ fraction of queries access the \spare (\cref{sec:prefix-ma}).
Space-wise, the expected number of fingerprints forwarded to the spare is at most a $1/\sqrt{2 \pi k}$ fraction of the fingerprints (\cref{sec:spare-analysis}).
This means that a bit consumed by the \spare adds an average of about $1/\sqrt{2 \pi k}$ bits of overhead to the overall filter's space requirement.
Our evaluation~(\cref{sec:evaluation}) shows that in practice, this translates to negligible space overhead.

\begin{figure}[t]
    \centering
    \includegraphics[width=0.7\columnwidth]{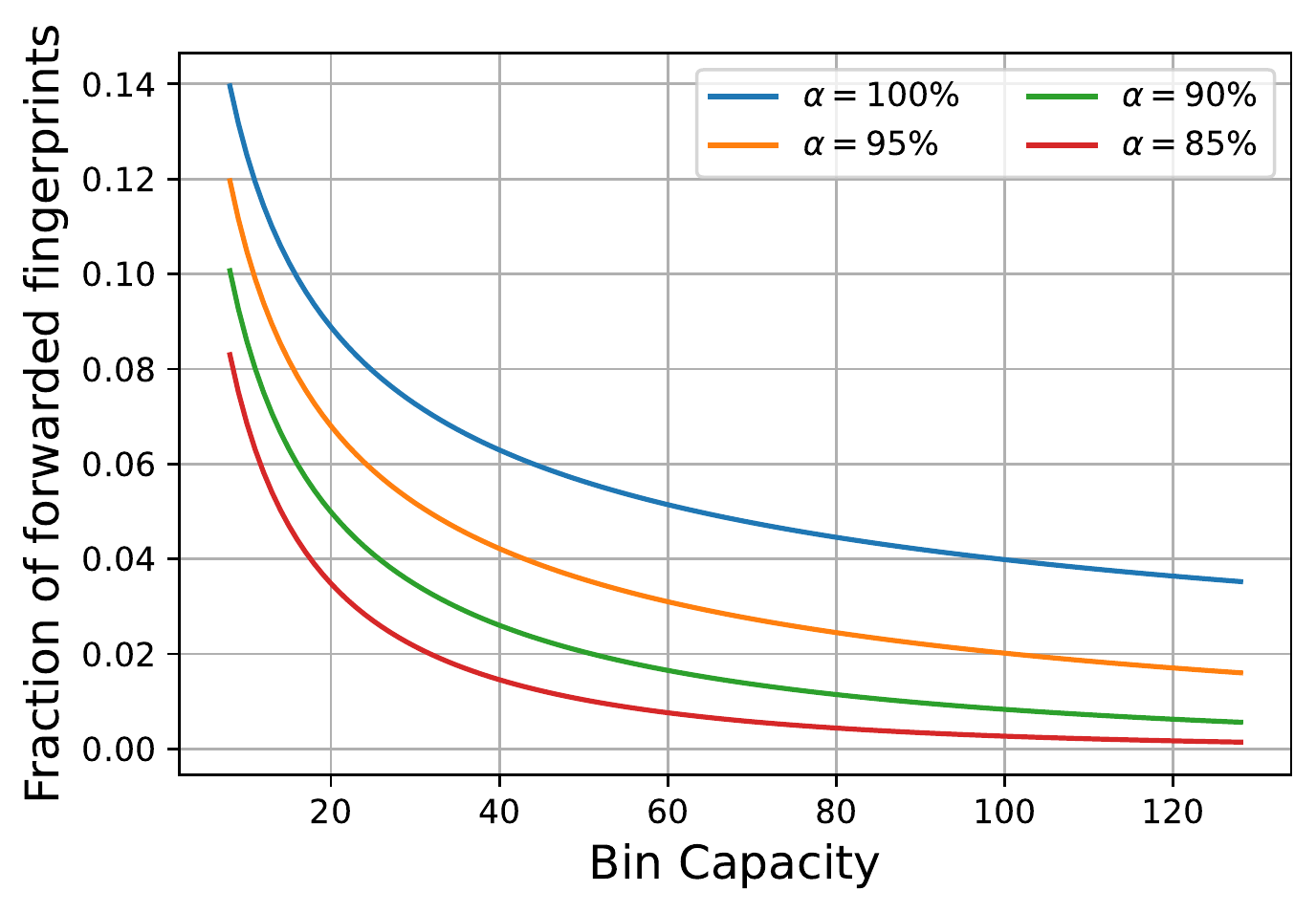}
    \caption{Expected fraction of fingerprints forwarded to the spare for $n=2^{30}$ as a function of the bin capacity $k$, for different values of the bin table maximal load factor ($\alpha$).
    }\label{fig:balls-in-spare}
\end{figure}

Importantly, however, the above formulas are ``worst case,'' in the sense that they are derived for a bin table of size $m=n/k$ (see~\cref{sec:analysis}).
In practice, we can forward significantly fewer fingerprints to the spare in exchange for a small amount of space by decreasing the bin table's maximal load factor, i.e., setting $m=\frac{n}{\alpha\cdot k}$ for $\alpha<1$, so the table's load factor after $n$ insertions is $\alpha$.
(Intuitively, with more bins, fewer bins overflow.) \cref{fig:balls-in-spare} shows this trade-off.
It plots the expected fraction of fingerprints forwarded to the spare for $n=2^{30}$ as a function of the bin capacity $k$ for different settings of the bin table's maximal load factor.

Our prototype implementation uses $k=25$, resulting in about $8\%$ of the dataset being stored in the \spare when the bin table size is $n/k$ ($\alpha=100\%$).
By simply picking $\alpha=95\%$, we can reduce this fraction by $1.36\times$ at a negligible space cost. %
We thus use $m=\frac{n}{0.95\cdot k}$ in our evaluation.

\subsection{Analysis Summary} \label{sec:the-pf:params}

We analyze the \PF's properties in~\cref{sec:analysis}. The following theorem summarizes our results:

\begin{theorem}[\PF]\label{thm:PF}
    Consider a prefix filter with a bin table of $m=\frac{n}{\alpha\cdot k}$ bins ($\alpha \leq 1$) and a \spare implementation whose space requirement is $S(n',\eps')$, where $n'$ and $\eps'$ are the \spare's dataset size and false positive rate, respectively. Let $n' = 1.1 \frac{n}{\sqrt{2\pi k}}$. Then:

    \begin{enumerate}[leftmargin=*]
        \item The filter's false positive rate is bounded by $\frac{\alpha\cdot k}{s} + \frac{1}{\sqrt{2\pi k}} \eps'$.
        \item For every sequence of operations containing at most $n$ insertions, the filter does not fail with probability $\geq 1-\cProbVar$.
        \item If the filter does not fail, then every query accesses a single cache line with probability $\geq 1-\cgam$.
              The fraction of insertions that access the \spare is at most $\frac{1.1}{\sqrt{2\pi k}}$ with probability $\geq 1-\cProbVar$.
        \item The filter uses
              $\frac{1}{\alpha} \cdot n \cdot\left( \log (s/k)+2\right) + S(n',\eps')$ bits of memory.
    \end{enumerate}
\end{theorem}
\begin{proof}
    Property (1) is proven in~\cref{sec:filter-fpp}.
    Property (2) is proven in~\cref{sec:spare-analysis}.
    Property (3) is proven in~\cref{sec:prefix-ma}.
    Property (4) follows by combining the space requirement of the \spare and the pocket dictionary bin implementation (\cref{sec:optimize}), which uses $k(\log (s/k)+2)$ bits per bin.
\end{proof}

\paragraph{Parameter and spare selection.}
\Cref{thm:PF} implies that the \PF can achieve a desired false positive rate $\eps$ in many ways, as long as $\frac{\alpha\cdot k}{s} + \frac{1}{\sqrt{2 \pi k}} \eps' \leq \eps$.
In practice, implementation efficiency considerations in our prototype dictate that $k/s=1/256$, which leads to a choice of $k=25$~(\cref{sec:pd-params}) mini-fingerprints per bin (each mini-fingerprint is in $[s]$).
This is similar to the situation in other practical filters, such as the cuckoo and vector quotient filters, where efficient implementations support only a few discrete false positive rates~\cite{dblp:conf/conext/fanakm14,pandey2021vector}.
Importantly, the \PF can obtain a false positive rate below $1/256$ by using $\alpha<1$.
Our prototype uses $\alpha=0.95$, which pays a negligible space cost to achieve $\eps < 1/256$ and also reduces the fraction of fingerprints stored in the \spare~(\cref{sec:the-pf:which-spare}).
The \spare's false positive rate is not a major factor in the filter's false positive rate, since it gets downweighted by a factor of $1/\sqrt{k}$.

\Cref{thm:PF} also implies that the \PF requires
$\frac{1}{\alpha}(\log (1/\eps)+2) + S(n',\eps')/n$ bits per key, assuming $k/s = \eps$.
The choice of \spare thus marginally affects the space requirement.
For instance, \cref{table:the-space} shows the space requirement obtained when using a \CF with $\eps' = \eps$ for a \spare.
This \spare consumes $S(n',\eps) = n' (\log(1/\eps) + 3)$ bits,
which for the overall filter translates to at most $\frac{1.1 \cdot (\log(1/\eps) + 3)}{\sqrt{2\pi k}}$ additional bits per key.

\paragraph{Lower failure probability for large datasets.}
The \PF's failure probability bound specified in~\cref{thm:PF} holds for all $n$, but it is not tight.
In~\cref{sec:analysis:failure} we show that for large $n$ (specifically, $n \geq 2^{28} \cdot k$) a tighter analysis obtains exponentially lower \PF failure probability bounds.
In particular, for our prototype PD implementation's choice of $k=25$, if the spare's dataset size is set to $n' = 1.015 \frac{n}{\sqrt{2\pi k}}$, then the \PF's filter probability is smaller than $2^{-40}$ for $n \geq 2^{28} \cdot k$.

\subsection{Discussion}

\paragraph{Concurrency support.}
The \PF admits a highly scalable concurrent implementation. Assuming a concurrent \spare implementation, the \PF needs only to synchronize accesses to the bin table.
Since \PF operations access one bin, this synchronization is simple to implement with fine-grained per-bin locking.
This scheme is simpler than in other load balancing schemes, such as cuckoo or power-of-two-choices hashing, which may need to hold locks on two bins simultaneously.

While a concurrent implementation and its evaluation is not in the scope of this paper, we expect it to be highly scalable.
Indeed, a similar (per-bin locking) scheme is used by the \VQF, whose throughput is shown to scale linearly with the number of threads~\cite{pandey2021vector}.

\paragraph{Fingerprint collisions.}
Our definition of an incremental filter requires all inserted keys to be distinct~(\cref{sec:back}), but there may exist distinct keys with identical fingerprints.
This event can pose a problem for certain ``hash table of fingerprints'' filter designs.
For example, a \CF with $b$-fingerprint buckets will fail after an insertion of $2b+1$ keys with the same fingerprint.

An incremental filter can avoid storing duplicate fingerprints by verifying (using a $\query$) that an inserted fingerprint is not already stored, but this solution slows down all insertions.
However, the \PF can cope with duplicate fingerprints in the following, more efficient, approach.
Instead of completely avoiding storing duplicate fingerprints, the \PF  only avoids storing duplicate fingerprints in the \spare.
That is, the \PF forwards a fingerprint to the \spare only if the fingerprint is not already present there (checked by a $\query$ of the \spare).
This approach slows down only the insertions forwarded to the spare (typically only $\frac{1.1}{\sqrt{2\pi k}}$ of all insertions~(\cref{sec:spare-analysis})) instead of slowing down all insertions.

Empirically, however, we find that when the \spare is implemented with dynamic filters such as the \CF or \VQF, storing duplicate fingerprints in the spare (i.e., avoiding the above check) does not cause the \spare to fail.
Therefore, our prototype \PF implementation does not check for duplicate fingerprints before forwarding an insertion to the \spare.

\paragraph{Comparison to the \BE.}
The \PF's two-level architecture is inspired by the theoretic dynamic \BE~\cite{DBLP:conf/swat/BerceaE20}.
The \PF does not need to support deletions, which allows us to change the \BE architecture in several important ways that are crucial for making the \PF a fast, practical filter:

\begin{enumerate}[leftmargin=*]
    \item \textbf{Eviction policy.} In the \BE, no evictions take place. A key is sent to the \spare if its bin is full upon insertion.
          As a result, a query has no way of knowing in which level its key may be stored, and must always search \emph{both} the bin table and the \spare.

    \item \textbf{Datatype of the \spare.} In the \BE, the \spare is a \emph{dictionary} of \emph{keys}.
          While a dictionary is not space efficient, the \BE's spare holds only $o(n / \log n)$ keys and its size is $o(n)$ bits~\cite{DBLP:conf/swat/BerceaE20}, i.e., asymptotically negligible. For practical values of both $n$ and $k$, however, the \spare stores about 6\%--8\% of the keys---which is far from negligible. The \PF solves this problem by implementing the \spare as a filter.

    \item \textbf{Universe of the \spare.} In the \BE, the \spare stores actual keys from $\UU$, which are forwarded if their bin is full upon insertion. In the \PF, the spare ``stores'' key fingerprints from the range of $\FP$, because forwarding (through eviction) can occur when the key itself is no longer available.
\end{enumerate}

\section{Pocket Dictionary Implementation}\label{sec:optimize}

The \PF's query speed is dictated by the speed of searching a first-level bin. In particular, a bin should fit into a single cache line, as otherwise queries may incur multiple cache misses. On the other hand,
we want to maximize the bin capacity $k$, as that improves load balancing, which affects the spare's size (\cref{sec:spare-analysis}) and the probability that negative queries need to search the \spare (\cref{sec:prefix-ma}).

To meet these conflicting demands, each first-level bin in the \PF is implemented with a \emph{pocket dictionary} (PD) data structure, introduced in the \BE~\cite{bercea2019fullydynamic}.
The PD is a space-efficient representation of a bounded-size set, which uses the Elias-Fano encoding~\cite{elias1974efficient,fano1971number} to encode the elements it stores~(see~\cref{sec:pd-background}).

The PD has constant-time operations, but in practice, all existing PD implementations~\cite{pandey2021vector,einziger2017tinyset} perform (relatively) heavy computations to search the compact encoding. In comparison, a cuckoo filter query---whose speed we would like to match or outperform---reads two memory words and performs simple bit manipulations on them, so computation is a negligible fraction of its execution time.

This section describes the \PF's novel PD implementation. Our implementation employs a \emph{query cutoff} optimization that exploits SIMD instructions to avoid executing heavy computations for $>99$\% of negative queries (\cref{sec:pd-opt}). We further extend the PD to support new operations required by the \PF (\cref{sec:pd-exts}).

Our PD implementation can be independently useful, as the PD (or a similar structure) is at the heart of filters such as the vector quotient filter (where PDs are called ``mini-filters''~\cite{pandey2021vector}) and TinySet~\cite{einziger2017tinyset} (where they are called ``blocks''). Indeed, we find that implementing the vector quotient filter using our PD outperforms the authors' original implementation, and therefore use our implementation in the evaluation (\cref{sec:evaluation}).

\subsection{Background: Pocket Dictionary}\label{sec:pd-background}

A pocket dictionary encodes a set of at most $k$ ``small'' \emph{elements} using a variant of the quotienting technique. In the \PF, each PD element is the mini-fingerprint of some \PF key.
Conceptually, a PD encodes $Q$ ``lists'' each of which contains $R$-bit values, subject to the constraint that at most $k$ values are stored overall. We use $PD(Q,R,k)$ to denote a concrete PD implementation with fixed values for these parameters.

Each PD element is a pair $(q,r) \in [Q] \times [2^R]$.
We refer to $q$ and $r$ as the \emph{quotient} and \emph{remainder}, respectively. The client of a $PD(Q,R,k)$ is responsible for ensuring that the most significant bits of elements can be viewed as a $q \in Q$, as $Q$ is not necessarily a power of two. In the \PF, our mini-fingerprint hash $\fp(\cdot)$ takes care of this.

\paragraph{Encoding.}
A $PD(Q,R,k)$ encodes $k$ elements using $R+2$ bits per element.
In general, a $PD(Q,R,k)$ represents $t \leq k$ elements with $Q + t + tR$ bits using the following encoding.
We think of the PD as having $Q$ lists, where inserting element $(q,r)$ to the PD results in storing $r$ in list $q$.
The $\PD$ consists of two parts, a $\hheader$ and $\bbody$.
The $\hheader$ encodes the occupancy of each list. These counts (including for empty lists) appear in quotient order, encoded in unary using the symbol $0$ and separated by the symbol $1$.
The $\bbody$ encodes the contents of the lists: the remainders of the PD's elements are stored as $r$-bit strings in non-decreasing order with respect to their quotient, thereby forming the (possibly empty) lists associated with each quotient.

Consider, for example, a $PD(8,4,7)$ storing the following set:
\begin{align*}
    \set{(1,13),(2,15),(3,3),(5,0),(5,5),(5,15),(7,6)}
\end{align*}

Then
\begin{align*}
    \hheader & \triangleq 1 \circ 01 \circ 01 \circ 01 \circ 1\circ0001\circ 1\circ 01 \\
    \bbody   & \triangleq 13 \circ 15 \circ 3 \circ 0 \circ 5 \circ 15 \circ 6,
\end{align*}

where the ``$\circ$'' symbol denotes concatenation and does not actually appear in the PD's encoding.

\paragraph{Operations.}
For $q \in Q$, we denote the occupancy of list $q$ by $occ(q)$,
and the sum of occupancies of all PD lists smaller than $q$ by $S_{q} = \sum_{q' < q} occ(q')$. PD operations are executed as follows.

\vspace{0.5\baselineskip}\noindent
\emph{$\query(q,r)$:} Compute $S_q$ and $occ(q)$.
Then, for every $S_q \leq i < S_q + occ(q)$, compare the input remainder $r$ with the remainder at $\bbody[i]$. If a match is found, return \yes; otherwise, return \no.

\vspace{0.5\baselineskip}\noindent
\emph{$\ins(q,r)$:} If the PD is full (contains $k$ remainders), the insertion fails.
Otherwise, the $\hheader$ is rebuilt by inserting a $0$ after the first $S_q + q$ bits.
Then, the $\bbody$ is rebuilt by moving the remainders of lists $q+1,\dots,Q$ (if any) one position up,
from $\bbody[j]$ to $\bbody[j+1]$, and inserting $r$ at $\bbody[S_q + occ(q)]$.

\paragraph{Implementation.}
Existing PD implementations use rank and select operations~\cite{SelectDef} to search the PD.
For a $b$-bit vector $B \in \{0,1\}^b$, $\Rank(B,j)$ returns the number of 1s in the prefix $B[0, \dots, j]$ of B and $\Select(B,j)$ returns the index of the $j$-th 1 bit. Formally:
\begin{align*}
    \Rank(B,j)   & = |\{ i \in \left[0 \dots j\right] \, | \, B[i] = 1 \}| \\
    \Select(B,j) & = \min \{0\leq i < b \, | \, \Rank(B,i) = j\},
\end{align*}
\noindent
where $\min\left( \emptyset\right)$ is defined to be $b$.

To perform a PD $\query(q,r)$, the implementation uses $\Select$ on the PD's $\hheader$ to retrieve the position of the $(q-1)$-th and $q$-th $1$ bits, i.e., the endpoints of the interval representing list $q$ in the $\bbody$. It then searches the $\bbody$ for $r$ in that range. If $R$ is small enough so that remainders can be represented as elements in an AVX/AVX-512 vector register, this search can be implemented with AVX vector instructions.
Unfortunately, implementing the $\Select$ primitive efficiently is challenging, with several proposals~\cite{pibiri2021rank,DBLP:journals/corr/MulaKL16,DBLP:journals/corr/PandeyBJ17},
none of which is fast enough for our context (see~\cref{sec:pd-find}).

\subsection{Optimized Pocket Dictionary} \label{sec:pd-opt}

Here we describe the \PF's efficient PD implementation. We explain the PD's physical layout and parameter choices (\cref{sec:pd-params}), the PD's search algorithm---its key novelty---that avoids executing a $\Select$ for $>99\%$ of random queries (\cref{sec:pd-find}), and PD extensions for supporting the \PF's insertion procedure (\cref{sec:pd-exts}).

\subsubsection{Physical Layout and Parameters} \label{sec:pd-params}

Our implementation has a fixed-sized header, capable of representing the PD's maximum capacity. Three considerations dictate our choice of the $Q$, $R$, and $k$ parameters:

\paragraph{\circled{1} Each PD in the \PF should reside in a single 64-byte cache line.}
This constraint guarantees that the first-level search of every \PF query incurs one cache miss. Satisfying this constraint effectively restricts possible PD sizes to 32 or 64 bytes, which are sizes that naturally fit within a 64-byte cache line when laid out in a cache line-aligned array. (Other sizes would require either padding the PDs---wasting space---or would result in some PDs straddling a cache line boundary, necessitating two cache misses to fetch the entire PD.)

\paragraph{\circled{2} The PD's header should fit in a 64-bit word.}
This constraint arises because our algorithm needs to perform bit operations on the header. A header that fits in a single word enables efficiently executing these operations with a handful of machine instructions.

\paragraph{\circled{3} The PD's body can be manipulated with SIMD (vector) instructions.} The AVX/AVX-512 instruction set extensions include instructions for manipulating the 256/512-bit vectors as vectors of 8, 16, 32, or 64-bit elements~\cite{intelAVXexts}. This means that R must be one of $\left\{8,16,32,64\right\}$.

\paragraph{Parameter choice.}
We choose parameters that maximize $k$ subject to the above constraints:
$Q=25$, $R=8$, and $k=25$. The $\PD(25, 8, 25)$ has a maximal size of $250 = 25 + 25 + 25\cdot8$ bits, which fit in 32 bytes with 6 bits to spare (for, e.g., maintaining PD overflow status), while having a maximal $\hheader$ size of $50 = 25+25$ bits, which fits in a machine word.

\subsubsection{Cutting Off Queries} \label{sec:pd-find}

Our main goal is to optimize negative queries (\cref{sec:back}).
For the \PF to be competitive with other filters, a PD query must complete in a few dozen CPU cycles. This budget is hard to meet with the standard PD search approach that performs multiple $\Select$s on the PD $\hheader$ (\cref{sec:pd-background}). The problem is that even the fastest x86 $\Select$ implementation we are aware of~\cite{DBLP:journals/corr/PandeyBJ17} takes a non-negligible fraction of a PD query's cycle budget, as it executes a pair of instructions (PDEP and TZCNT) whose latency is 3--8 CPU cycles each~\cite{abel2019uops}.

We design a PD search algorithm that executes a negative query without computing $\Select$ for $>99$\% of queries, assuming uniformly random PD elements---which is justified by the fact that our elements are actually mini-fingerprints, i.e., the results of hashing keys.

Our starting point is the observation that most negative queries can be answered without searching the $\hheader$ at all; instead, we can search the $\bbody$ for the remainder of the queried element, and answer \no if it is not found. Given a queried element $(q,r)$, searching the $\bbody$ for a remainder $r$ can be implemented efficiently using two AVX/AVX-512 vector instructions: we first use a ``broadcast'' instruction to create a vector $x$ where $x[i]=r$ for all $i \in [k]$, and then compare $x$ to the PD's $\bbody$ with a vector compare instruction. The result is a word containing a bitvector $\mathtt{v}_r$, defined as follows
\begin{align*}
    \forall i\in[k]\quad \mathtt{v}_r[i] & \triangleq
    \begin{cases}
        1 & \text{if}\quad r = \bbody[i] \\
        0 & \text{otherwise}
    \end{cases}
\end{align*}

\noindent
Then if $\mathtt{v}_r = 0$, the search's answers is \no.

For the \PF's choice of PD parameters, the above ``cutoff'' can answer \no for at least 90\% of the queries (\cref{lemma:shortcut} below).
The question is how to handle the non-negligible~$\approx 10$\% of queries for which the ``cutoff'' cannot immediately answer \no, without
falling back to the standard PD search algorithm that performs multiple $\Select$s.

Our insight is that in the vast majority of cases where a query $(q,r)$ has $\mathtt{v}_r \neq 0$, then $r$ appears \emph{once} in the PD's $\bbody$ (\cref{lemma:shortcut2} below). Our algorithm therefore handles this common case without executing $\Select$, and falls back to the $\Select$-based solution only in the rare cases where $r$ appears more than once in the PD's $\bbody$.

\Cref{algo:pop1} shows the pseudo code of the \PF's PD search algorithm.
The idea is that if $\bbody[i]=r$ ($0 \leq i < k$), we can check if index  $i$ belongs to list $q$ by verifying in the $\hheader$ that (1) list $q-1$ ends before index $i$ and (2) list $q$ ends after index $i$. We can establish these conditions by checking that $\Rank(\hheader, q+i-1)=q$, which implies (1), and that list $q$ is not empty, which implies (2).
For example, suppose $q=4$, $i=3$, and $\hheader = 101010110001101$. Then list 3 ends before index 3 but list 4 is empty (bit $q+i$ is 1).

Computing $\Rank$ costs a single-cycle POPCOUNT instruction. We further
optimize by avoiding explicitly extracting $i$ from $\mathtt{v}_r$; instead, we leverage the fact that the only set bit in $\mathtt{v}_r$ is bit $i$, so $(\mathtt{v}_r \ll q)-1$ is a bitvector of $q+i$ ones, which we bitwise-AND with the $\hheader$ to keep only the relevant bits for our computation.

\begin{algorithm}[t]
    \setcounter{AlgoLine}{0}
    \KwIn {Queried element $(q,r)$.}
    \KwOut{Whether $(q,r)$ is in the PD.}
    construct 64-bit bitvector $\mathtt{v}_r$; \Comment{\small Using VPBROADCAST \& VPCMP}
    \If{$\mathtt{v}_r = 0$}{\Return \no }
    \eIf(\Comment*[f]{\small $\mathtt{v}_r$ has one set bit?}){$(\mathtt{v}_r\,\&\,(\mathtt{v}_r-1)) = 0$} {$w \gets \mathtt{v}_r \ll q$ \;
        \eIf{ $\Rank(\hheader\,\&\,(w-1), 64) = q$ \textbf{and} $(\hheader\,\&\, w) = 0 $ }{ \Return{ \yes } }{ \Return \no }
    }
    {
        use $\Select$-based algorithm \;
    }
    \caption{PD search algorithm.}\label{algo:pop1}
\end{algorithm}

\paragraph{Analysis.}
We prove that if PD and query elements are uniformly random, then a query $(q,r)$ has $\mathtt{v}_r=0$ with probability~$>0.9$ (\cref{lemma:shortcut}), and if $\mathtt{v}_r \neq 0$, then with probability~$>0.95$, $r$ will appear once in the PD's body (\cref{lemma:shortcut2}).

\begin{claim}
    \label{lemma:shortcut}
    Consider a $\PD(25, 8, 25)$ containing uniformly random elements. Then the probability over queries $(q,r)$ that $\mathtt{v}_r=0$ is $>0.9$.
\end{claim}
\begin{proof}
    Consider a $\PD(Q, R, k)$.
    Let the number of distinct remainders in the PD's $\bbody$ be $s \leq k$. For a uniformly random remainder $r$,
    \begin{align*}
        \Pr{\mathtt{v}_r = 0} = 1 - \Pr{\mathtt{v}_r \neq 0} & = 1 - s/2^R \geq 1 - k/2^R
    \end{align*}
    For the \PF's choice of PD parameters, we have $1 - k/2^R \overset{k = 25,R=8}{\approx} 0.902$.
\end{proof}

\begin{claim} \label{lemma:shortcut2}
    Consider an element $(q,r)$. The probability (over $\PD(25, 8, 25)$ with uniformly random elements) that a PD's $\bbody$ contains $r$ once, conditioned on $\mathtt{v}_r \neq 0$, is~$>0.95$.
\end{claim}
\begin{proof}
    Consider a $\PD(Q,R,k)$ whose body contains $s \leq k$ distinct remainders.
    The probability that the PD's body contains $r$ once, conditioned on it containing $r$ at all, is
    \begin{align*}
        \frac{s \cdot 1/2^R \cdot (1-1/2^R)^{s-1}}{1-(1-1/2^R)^s} \geq
        \frac{k \cdot 1/2^R \cdot (1-1/2^R)^{k-1}}{1-(1-1/2^R)^k} \overset{k = 25,R=8}{\approx} 0.953
    \end{align*}

    The first inequality holds because for $\beta\in (1/2,1)$ and $i \geq 1$, $T(i) \triangleq \frac{i \cdot \beta^{i-1}}{1-\beta^i}$ is monotonically decreasing; in our case, $\beta = 1-1/2^R$. The claim then follows since the bound holds for any $1 \leq s \leq k$.
\end{proof}

\subsubsection{Prefix Filter Support: Finding the Maximum Element} \label{sec:pd-exts}

A \PF insertion that tries to insert an element into a full PD must locate (and possibly evict) the maximum element stored in the PD. Here, we describe the extensions to the PD algorithm required to support this functionality.

\paragraph{Computing the remainder.}
In a standard PD~(\cref{sec:pd-background}), the $\bbody$ stores the remainders according to the lexicographic order of the elements.
Maintaining this ``lexicographic invariant'' is wasteful, hence we use the following relaxed invariant: if the PD has overflowed, then the remainder of the maximum element is stored in the last ($k$-th) position of the $\bbody$.
Maintaining this relaxed invariant requires finding the maximum remainder in the last non-empty list only when the PD first overflows or when the the maximum element is evicted.

\paragraph{Computing the quotient.}
In a standard PD, the maximum element's quotient needs to be computed from the $\hheader$: it is equal to the quotient of the last non-empty list. This quotient can be derived from the number of ones before the $k$-th zero bit in the $\hheader$ or from the number of consecutive (trailing) ones in the $\hheader$.
We find, however, that computing these numbers adds non-negligible overhead for a filter operation. To avoid this overhead, the \PF's PD reserves a $\log Q$-bit field in the PD which stores (for a PD that has overflowed) the quotient of the maximum element stored in the PD.

\section{Analysis} \label{sec:analysis}

In this section, we analyze the properties of the prefix filter.  The
analysis holds for every sequence of query and insertion operations
provided that there are at most $n$ insertions.
We assume that the hash functions are random functions (i.e., the function values are uniformly distributed and independent).
The analysis assumes the following order:
(1)~A sequence $\sigma$ of operations (consisting of queries and at most $n$ insertions of distinct keys) is chosen and fixed.
(2)~A random hash function $h$ is chosen.
(3)~The filter employs the hash function $h$ to execute the sequence of operations $\sigma$.
In particular, the sequence $\sigma$ cannot be changed based on results obtained for earlier queries (e.g., false positive responses).
Hence, the  underlying probability space is induced only by the random choice of the hash function.  The justification for such an analysis in practice is based on the assumption that the sequence of operations is generated without knowledge of failure or false positive events.

\paragraph{Notation.}
\begin{table}[!htbp]
    \begin{tabular}{lp{0.6\columnwidth}}
        \toprule
        $n$                                              & the maximum number of elements in the set $\DD\subset \UU$.                                 \\
        $k$                                              & the capacity of each bin (maximum number of mini-fingerprints that can be stored in a bin). \\
        $m \triangleq n/k$                               & the number of bins in the bin table (i.e., first level).                                    \\
        $\FP(x)=(\bin(x),\fp(x))$                        & the fingerprint $\FP(x)$ of $x\in\UU$ is the pair $(\bin(x),\fp(x))\in [m]\times [s]$.
        \\
        \midrule
        $p\triangleq 1/m$                                & the probability that a ball falls in a specific bin.                                        \\
        $B_i$                                            & the number of balls (i.e., elements in $\DD$) mapped to the $i$'th bin.                     \\
        $X_i=\max\{B_i - k, 0\}$                         & the number of balls sent to \spare from the $i$'th bin.                                     \\
        $X=                     \sum_{1\leq i\leq m}X_i$ & the number of balls sent to the \spare.                                                     \\
        $\Bin{n}{p}$                                     & a binomial random variable with parameters $n$ and $p$.                                     \\
        \bottomrule
    \end{tabular}
    \caption{Notation for the analysis. Top: Filter notation. Bottom: Notation for ``balls into bin'' analysis.
    }\label{table:not-balls-in-bins}
\end{table}

We use the following notation (top of~\cref{table:not-balls-in-bins}):
\begin{enumerate*}[label=(\arabic*)]
    \item $\DD$ denotes the data set, and $n$ denotes the maximum number of elements in $\DD$;
    \item $m$ denotes the number of bins in the bin table (i.e., first level) and $p\triangleq 1/m$;
    \item $k$ denotes the capacity of each bin (which is an integer); and
    \item fingerprints are pairs $\FP(x)=(\bin(x),\fp(x))$, where $\bin(x)\in[m]$ and $\fp(x)\in [s]$.
    We refer to  $\fp(x)$ as the \emph{mini-fingerprint} of $x$.
\end{enumerate*}

\subsection{Spare Occupancy and Failure}\label{sec:spare-analysis}
In this section, we bound the number of elements that are forwarded to the \spare.
The analysis views the problem as a balls-into-bins experiment.
Namely, randomly throw $n$ balls (fingerprints) into $m=n/k$ bins of capacity $k$, such that if a ball is mapped to a full bin, then it is forwarded to the \spare.%
\footnote{Increasing the number of bins (i.e., $m > n/k$) does not increase the number of balls forwarded to the spare.}
Previous analyses of two-level filters require $m=(1+o(1))n/k$,
where the term $o(1)$ is greater than $1$ even for $n=2^{40}$~\cite{arbitman2010backyard,bercea2020dynamic}.

Let $B_i$ denote the number of balls that are mapped to bin $i$.
Let $\Bin{n}{p}$ denote a binomial random variable with parameters $n$ and $p$.
The random variables $B_i$ and $\Bin{n}{p}$ are identically distributed.
The contribution of bin $i$ to the spare is $X_i \triangleq\max\set{0,B_i-k}$.
Let $X\triangleq \sum_{i=1}^m X_i$ denote the random variable that equals the number of fingerprints that are forwarded to the spare.
\Cref{table:not-balls-in-bins} (bottom) summarizes these definitions.

We prove bounds on $X$ for practical values of $n$ (i.e., $n\geq 2^{25}$) using second moment bounds (Cantelli's inequality), as well as concentration bounds based on Hoeffding's inequality.
The former is better for small values of $n$ (see~\cref{sec:analysis:failure}).

\begin{theorem}\label{thm:X}
    The number $X$ of balls forwarded to the spare satisfies:
    \begin{align}
        \label{eq:expectation}
         & \expectation{}{X}=n\cdot(1-p)\cdot \Pr{\Bin{n}{p}=k}\leq n\cdot \frac{1}{\sqrt{2\pi k}} \\ &
        \label{eq:deviation}
        \Pr{X \geq (1+\delta) \cdot \expectation{}{X}} \leq \min\set{\frac{2\pi k}{\delta^2\cdot 0.99 n}, \exp\brak{\frac{-\delta^2 m\cdot 0.99 (1-p)}{\pi k}}}\;\;\text{(for $n\geq 5k, k\geq 20$)}
    \end{align}
\end{theorem}
\begin{proof}[Proof of \Cref{eq:expectation}]
    We prove the following propositions:
    \begin{align*}
        \mathbb{E}[X_i]                          & =k\cdot \Pr{\Bin{n}{p}\leq k} -
        \sum_{j=0}^{k}j\cdot \Pr{\Bin{n}{p} = j} &                                                             & (\Cref{prop:holes})                                   \\
        \sum_{j=0}^{k}\;                         & j\cdot\Pr{\Bin{n}{p} = j}= k\cdot \Pr{\Bin{n-1}{p}\leq k-1} &                     & (\Cref{prop:binomial identity}) \\
        \mathbb{E}[X_i]                          & =(1-p)\cdot k\cdot \Pr{\Bin{n}{p}= k}                       &                     & (\Cref{prop:bound Xi})
    \end{align*}
    We then use Stirling's approximation to bound $\Pr{\Bin{n}{p}= k}$ by $\frac{1}{\sqrt{2\pi k (1-p)}}$  (\Cref{prop:mu}) and finally obtain~\Cref{eq:expectation} by linearity of expectation, as $X = \sum_{i=1}^m X_i$ and $m=n/k$.

    \begin{proposition}[$\Exp{X_i}$]\label{prop:holes} For every bin $i$,
        \begin{align*}
            \mathbb{E}[X_i] & =k\cdot \Pr{\Bin{n}{p}\leq k} -
            \sum_{j=0}^{k}j\cdot \Pr{\Bin{n}{p} = j}  \;.
        \end{align*}
    \end{proposition}
    \begin{proof}
        \renewcommand{\qedsymbol}{\ensuremath{\blacksquare}}
        Because $\sum_i B_i=n=km$, the number of
        vacant slots in the bins also equals $X$.
        Namely,
        $X=\sum_{i: B_i\geq k} (B_i-k)=
            \sum_{i: B_i\leq k} (k-B_i)$.
        Let $I_{B_i\leq k}$ denote the
        indicator variable that equals $1$ iff $B_i\leq k$. By taking the
        expectation from both sides, we obtain
        $\expectation{}{X}=\expectation{}{\sum_{i=1}^m (k-B_i)\cdot
                I_{B_i\leq k}}$.
        The random variables $\set{X_i}$ are identically distributed, hence
        it suffices to prove the proposition for $X_1$.
        Because $\expectation{}{X}=m\expectation{}{X_1}$, by linearity of expectation,
        \begin{align*}
            \expectation{}{X_1} & = m^{-1} \sum_{i=1}^m \expectation{}{(k-B_i)\cdot I_{B_i\leq k}} \\
                                & =\expectation{}{(k-B_1)\cdot I_{B_1\leq k}}                      \\
                                & =
            k\cdot \Pr{B_1\leq k} -
            \sum_{j=0}^{k}j\cdot \Pr{B_1 = j} \;.
        \end{align*}
        The proposition follows, as $B_1$ is distributed binomially with parameters $n$ and $p$.
    \end{proof}

    \begin{proposition}[Truncated binomial expectation]\label{prop:binomial identity}
        \begin{align*}
            \sum_{j=0}^{k}j\cdot\Pr{\Bin{n}{p} = j} & = k\cdot \Pr{\Bin{n-1}{p}\leq k-1}
        \end{align*}
    \end{proposition}
    \begin{proof}
        \renewcommand{\qedsymbol}{\ensuremath{\blacksquare}}
        The second line below uses the identity
        $j{n\choose j} = n{n-1\choose j-1}$.
        \begin{align*}
            \sum_{j=0}^{k}\Pr{\Bin{n}{p} = j}\cdot j & =
            \sum_{j=0}^{k}{n\choose j}p^j \cdot\brak{1-p}^{n-j} \cdot j           \\
                                                     & =
            \sum_{j=1}^{k}n\cdot{n-1\choose j-1}p^j \cdot\brak{1-p}^{n-j}         \\ &=
            np\sum_{j=1}^{k}{n-1\choose j-1}p^{j-1} \cdot\brak{1-p}^{(n-1)-(j-1)} \\ &=
            k\sum_{j=0}^{k-1}{n-1\choose j}p^{j} \cdot\brak{1-p}^{n-1-j}          \\ &=
            k\cdot \Pr{\Bin{n-1}{p} \leq k-1}
        \end{align*}
    \end{proof}

    \begin{proposition}\label{prop:bound Xi}
        \begin{align*}
            \expectation{}{X_i} & =(1-p) \cdot k\cdot \Pr{\Bin{n}{p}= k}
        \end{align*}
    \end{proposition}
    \begin{proof}
        \renewcommand{\qedsymbol}{\ensuremath{\blacksquare}}
        By~\cref{prop:holes} and~\cref{prop:binomial identity},
        \begin{align*}
            \expectation{}{X_i} & =k\cdot \Pr{\Bin{n}{p}\leq k}-k\cdot\Pr{\Bin{n-1}{p}\leq k-1}\;.
        \end{align*}
        Let $Y_n$ denote the last trial in the binomial experiment (i.e.,
        $\Pr{Y_n=1}=p$). The event $(\Bin{n}{p}\leq k)$ is the union of two disjoint events:
        \begin{align*}
            \brak{\Bin{n}{p}\leq k} & = \brak{\Bin{n-1}{p}\leq (k-1)}
            \sqcup
            \brak{(\Bin{n-1}{p}=k) \land (Y_n=0)}\;.
        \end{align*}
        Hence,
        \begin{align*}
            \Pr{\Bin{n}{p}\leq k}-\Pr{\Bin{n-1}{p}\leq k-1} & =
            \Pr{\Bin{n-1}{p}=k}\cdot (1-p)\;.
        \end{align*}

        To complete the proof of the proposition, we show that if $p = k/n$,        it holds that $\Pr{\Bin{n-1}{p}=k} = \Pr{\Bin{n}{p}=k}$:
        \begin{align*}
            \Pr{\Bin{n-1}{p}=k} & = \binom{n-1}{k} p^{k} (1-p)^{n-1-k}         \\ &=
            \overbrace{\frac{n-k}{n}}^{=(1-p)}\binom{n}{k} p^{k} (1-p)^{n-1-k} \\&=
            \binom{n}{k} p^{k} (1-p)^{n-k}                                     \\&= \Pr{\Bin{n}{p}=k}
        \end{align*}
    \end{proof}
    \begin{proposition}[Tight Bounds on $\Pr{\Bin{n}{p}=k}$]\label{prop:mu}
        \begin{align*} %
            \frac{\exp(t_0)}{\sqrt{2\pi k \cdot (1-p)}} < \Pr{\Bin{n}{p}=k} < \frac{\exp(t_1)}{\sqrt{2\pi k \cdot (1-p)}}
        \end{align*}
        where
        \begin{align*}
            t_0 & = \frac{1}{12n + 1} - \brak{\frac{1}{12k} + \frac{1}{12(n-k)}}     \\
            t_1 & = \frac{1}{12n} - \brak{\frac{1}{12k + 1} + \frac{1}{12(n-k) + 1}} \\
        \end{align*}
    \end{proposition}
    \begin{proof}
        \renewcommand{\qedsymbol}{\ensuremath{\blacksquare}}
        We use the following notation:
        \begin{align*}
            \alpha(n) \triangleq & \sqrt{2 \pi n}\ \brak{\frac{n}{e}}^n e^{\frac{1}{12n + 1}} \\
            \beta(n) \triangleq  & \sqrt{2 \pi n}\ \brak{\frac{n}{e}}^n e^{\frac{1}{12n}}
        \end{align*}

        We first prove the upper bound. The proof is based on replacing the binomial coefficient with Stirling's approximation, i.e.,
        \begin{align*}
            \alpha(n) = \sqrt{2 \pi n}\ \brak{\frac{n}{e}}^n e^{\frac{1}{12n + 1}} < n! < \sqrt{2 \pi n}\ \brak{\frac{n}{e}}^n e^{\frac{1}{12n}} = \beta(n)
        \end{align*}
        \begin{align*}
            \Pr{\Bin{n}{p}=k} & = \binom{n}{k} p^{k} (1-p)^{n-k}                                                                                                                                                                                                                                         \\\leq &
            \frac{\beta(n)}{\alpha(n-k)\alpha(k)}\cdot p^{k} (1-p)^{n-k}                                                                                                                                                                                                                                 \\    =&
            \frac{1}{\sqrt{2\pi}}\cdot \underbrace{\frac{\sqrt{n}}{\sqrt{(n-k)k}}}_{=\frac{1}{\sqrt{k(1-p)}}}\frac{\brak{\frac{n}{e}}^n \cdot \exp\brak{\frac1{12n}}}{\brak{\frac{n-k}{e}}^{n-k}\brak{\frac{k}{e}}^k \cdot \exp\brak{\frac{1}{12(n-k) + 1} + \frac{1}{12k + 1}}} \cdot p^{k} (1-p)^{n-k} \\     &%
            =
            \frac{1}{\sqrt{2\pi k (1-p)}} \cdot \brak{\frac{1}{1-p}}^{n-k}\brak{\frac{1}{p}}^{k} \cdot \exp(t_1)\cdot p^{k} (1-p)^{n-k} =
            \frac{\exp(t_1)}{\sqrt{2\pi k (1-p)}}
        \end{align*}
        This completes the upper bound on $\Pr{\Bin{n}{p}=k}$.

        Next, we prove the lower bound.
        The proof is analogous to the upper bound proof, with the difference of ``switching'' between the $\alpha$'s and the $\beta$'s:
        \begin{align*}
            \binom{n}{k} p^{k} (1-p)^{n-k} \geq \frac{\alpha(n)}{\beta(n-k)\beta(k)}\cdot p^{k} (1-p)^{n-k}
        \end{align*}
        and after rearranging things in the same manner, we get
        \begin{align*}
            \Pr{\Bin{n}{p}=k}\geq \frac{\exp(t_0)}{\sqrt{2\pi k (1-p)}}
        \end{align*}
    \end{proof}

    Observe that for $k\geq20$ and $n \geq 5k$,
    \begin{align*}
        1 & > \exp(t_1) > \exp(t_0) > \exp(2\cdot t_0) > 0.99,
    \end{align*}
    \Cref{eq:expectation} follows from this inequality and~\cref{prop:bound Xi,prop:mu}.
\end{proof}

\begin{proof}[Proof of \Cref{eq:deviation}]
    We bound $\Pr{X \geq (1+\delta) \cdot \expectation{}{X}}$ using the following propositions:
    \begin{align*}
        \Pr{X \geq (1+\delta) \cdot \expectation{}{X}}         & \leq \frac{2\pi k}{\delta^2\cdot 0.99 n} & (\cref{prop:cantelli}) \\
        \Pr{X \geq (1+\delta) \cdot \expectation{}{X}}         & \leq
        \exp\brak{\frac{-\delta^2 m\cdot 0.99 (1-p)}{\pi k}} & (\cref{prop:hoef})
    \end{align*}
    under the assumption that $n\geq 5k, k\geq 20$.
    \begin{proposition}\label{prop:cantelli}
        \begin{align*}
            \Pr{X \geq (1+\delta) \cdot \expectation{}{X}} \leq \frac{2\pi k}{\delta^2\cdot 0.99 n}
        \end{align*}
    \end{proposition}
    \begin{proof}
        \renewcommand{\qedsymbol}{\ensuremath{\blacksquare}}
        We prove~\Cref{prop:cantelli} using Cantelli's inequality, restated below:

        \begin{theorem}[Cantelli's Inequality]\label{ineq:cantelli}
            \begin{align*}
                \Pr{X \geq (1+\delta)\Exp{X}} \leq \frac{\variance{X}}{\variance{X}+\delta^2 \Exp{X}^2}.
            \end{align*}
        \end{theorem}

        Our proof relies on the following claim, which is proven later:
        \begin{claim}\label{claim:var-main}
            $\begin{aligned}
                    \variance{X}\leq m\cdot \variance{B_i}
                \end{aligned}$.
        \end{claim}

        Now, note that $\variance{B_i}= m\cdot n \cdot p(1-p) = n(1-p)$.
        Also,
        For $n\geq 5k$ and $k\geq 20$, we have
        \begin{align*}
            \expectation{}{X} \overset{\Cref{eq:expectation}}{=} n\cdot (1-p) \Pr{\Bin{n}{p}=k} \overset{\cref{prop:mu}}{\geq} n\cdot \sqrt{\frac{1-p}{2\pi k}}\cdot \exp(t_0) \overset{\substack{k \geq 20 \\ n\geq 5k}}{\geq} n\cdot \sqrt{\frac{0.99(1-p)}{2\pi k}}
        \end{align*}

        Because the function $f(v)=v/(v+a)$ is increasing for $v>0$ and $a>0$, the above imply the following:

        \begin{align*}
            \frac{\variance{X}}{\variance{X}+\delta^2 \expectation{}{X}^2} \leq
            \frac{n(1-p)}{n(1-p)+\delta^2 n^2\cdot {\frac{0.99(1-p)}{2\pi k}}} =
            \frac{1}{1+\delta^2 n\cdot {\frac{0.99}{2\pi k}}} \leq
            \frac{2\pi k}{\delta^2\cdot0.99 n}.
        \end{align*}
        Which completes the proof of~\Cref{prop:cantelli}.
        We are therefore left with proving~\Cref{claim:var-main}.
        \begin{proof}[Proof of~\Cref{claim:var-main}]
            \renewcommand{\qedsymbol}{\ensuremath{\blacksquare}}
            We will show that
            \begin{align*}
                \variance{X_i} & \leq \variance{B_i}       \\%\label{ineq:var1}\\
                \variance{X}   & \leq m\cdot\variance{X_i} %
            \end{align*}
            For the former inequality, note that $X_i^2\leq (B_i-k)^2$.
            Hence,
            \begin{align*}
                \variance{X_i}\leq \expectation{}{X_i^2} \leq
                \expectation{}{(B_i-k)^2}=\variance{B_i}\;.
            \end{align*}
            To prove the latter inequality note that
            the random variables $\set{B_i}$ are negatively associated~\cite[Theorem 13]{dubhashi1998balls} (which we denote by N.A.) and hence, their covariance is non-positive.
            Because $X_i$ is a monotone function of $B_i$, it follows that the random variables $\set{X_i}$ are also negatively associated~\cite{dubhashi1998balls}.
            Hence,
            \begin{align*}
                \variance{X}\overset{N.A}{\leq} m\cdot \variance{X_1} \leq m\cdot \variance{B_1} = m \cdot n \cdot p(1-p)=n(1-p)\;.
            \end{align*}
        \end{proof}
        This completes the proof of~\Cref{prop:cantelli}.
    \end{proof}

    \begin{proposition}\label{prop:hoef}
        \begin{align*}
            \Pr{X \geq (1+\delta) \cdot \expectation{}{X}} \leq \exp\brak{\frac{-\delta^2 m\cdot 0.99 (1-p)}{\pi k}}
        \end{align*}
    \end{proposition}
    \begin{proof}
        \renewcommand{\qedsymbol}{\ensuremath{\blacksquare}}

        Our proof relies on Hoeffding's bound~\cite{hoeffding1994probability}, restated below:
        \begin{theorem}[Hoeffding's Bound]\label{thm:hoef reff}
            Let $(X_1,\dots,X_m)$ be independent random variables such that $X_i\in[a,b]$. Consider the sum of these random variables, $X=\sum_{i=1}^m X_i$.
            Then %
            for all $\delta \geq 0$,
            \begin{align*}
        \Pr{X-\Exp{X}\geq \delta\cdot \Exp{X}}\leq \exp\brak{-\frac{2\brak{\delta\cdot \Exp{X}}^2}{m(b-a)^2}}
            \end{align*}
        \end{theorem}
        \begin{remark}
            Hoeffding's bound also applies to variables that are not independent provided they are negatively associated \cite[Proposition 5]{dubhashi1998balls}.
        \end{remark}

        We use Hoeffding's bound to prove the following exponential bound on $X-\Exp{X}$:
        \begin{align*}
        \Pr{X > (1+\delta) \cdot \Exp{X}}\leq \exp\brak{-2 \frac{\brak{\delta\cdot \Exp{X}}^2}{m\cdot k^2}}
        \end{align*}

        Define $Y_i\triangleq \max\set{0, k-B_i}$.
        Because $\sum_i B_i=n=km$, the number of vacant slots in all bins equals the number of balls that were forwarded to the \spare.    Therefore, we have
        \begin{align*}
            X = \sum_{i=1}^m X_i = \sum_{i=1}^m Y_i
        \end{align*}
        We use Hoeffding's bound on the sum of $\set{Y_i}_{i=1}^m$, because using Hoeffding's bound directly on the sum of $\set{X_i}_{i=1}^m$ would yield a weaker bound, as the ${\set{Y_i}_{i=1}^m}$ support is contained in an interval of length $k$, whereas the ${\set{X_i}_{i=1}^m}$ support might not be.\footnote{For example, if $k=O(1)$ and $n$ sufficiently large.}

        Recall that ${\set{B_i}_{i=1}^m}$ are negatively associated~\cite[Theorem 13]{dubhashi1998balls}.
        The ${\set{Y_i}_{i=1}^m}$ are also negatively associated, because
        each $Y_i$ satisfies $Y_i = \max\set{0, k-B_i}$, and therefore is an
        image of monotonically non-increasing function of $B_i$, which (by~\cite[Proposition 7(2)]{dubhashi1998balls}) implies that the $\set{Y_i}_{i=1}^m$ are also negatively associated. We can therefore use Hoeffding's bound on the $\set{Y_i}_{i=1}^m$.
    We apply the bound with the parameters $a=0, b = k$, proving the proposition:
        \begin{align*}
        \Pr{X > (1+\delta)\Exp{X}} & \leq
        \exp\brak{-2\frac{\brak{\delta\cdot \Exp{X}}^2}{m\cdot k^2}}
        = \exp\brak{-2\frac{\brak{\delta\cdot n(1-p)\Pr{\Bin{n}{p}=k}}^2}{m\cdot k^2}}\\
        & \leq
        \exp\brak{-\delta^2 m \frac{(1-p)\cdot \exp(2 t_0)}{\pi k}}
        \overset{\substack{k \geq 20 \\ n \geq 5k}}{\leq}
        \exp\brak{-\delta^2 m \frac{0.99(1-p)}{\pi k}},
    \end{align*}
    where the equality holds because $\Exp{X} = n(1-p)\Pr{\Bin{n}{p}=k}$ (by~\Cref{eq:expectation}) and $n=m \cdot k$, and the subsequent inequality follows from~\cref{prop:mu}.

    \end{proof}
    This completes the proof of~\Cref{eq:deviation} and thus of~\cref{thm:X}.
\end{proof}

\subsubsection{Failure Probability} \label{sec:analysis:failure}
In~\cref{sec:the-pf:spare-params}, we propose to set the \spare's dataset size to $1.1 \cdot \expectation{}{X}$.
Plugging in the $\delta=0.1$ implied by this setting into~\Cref{eq:deviation} yields
\begin{claim}\label{claim:fail}
    If the \spare's dataset size $n'$ is set to be $1.1 \cdot \expectation{}{X}$, then the probability that the \spare overflows (and hence the prefix filter fails) is at most $\cProbVar$.
\end{claim}

\begin{figure*}[b]
    \centering
    \includegraphics[width=0.85\textwidth]{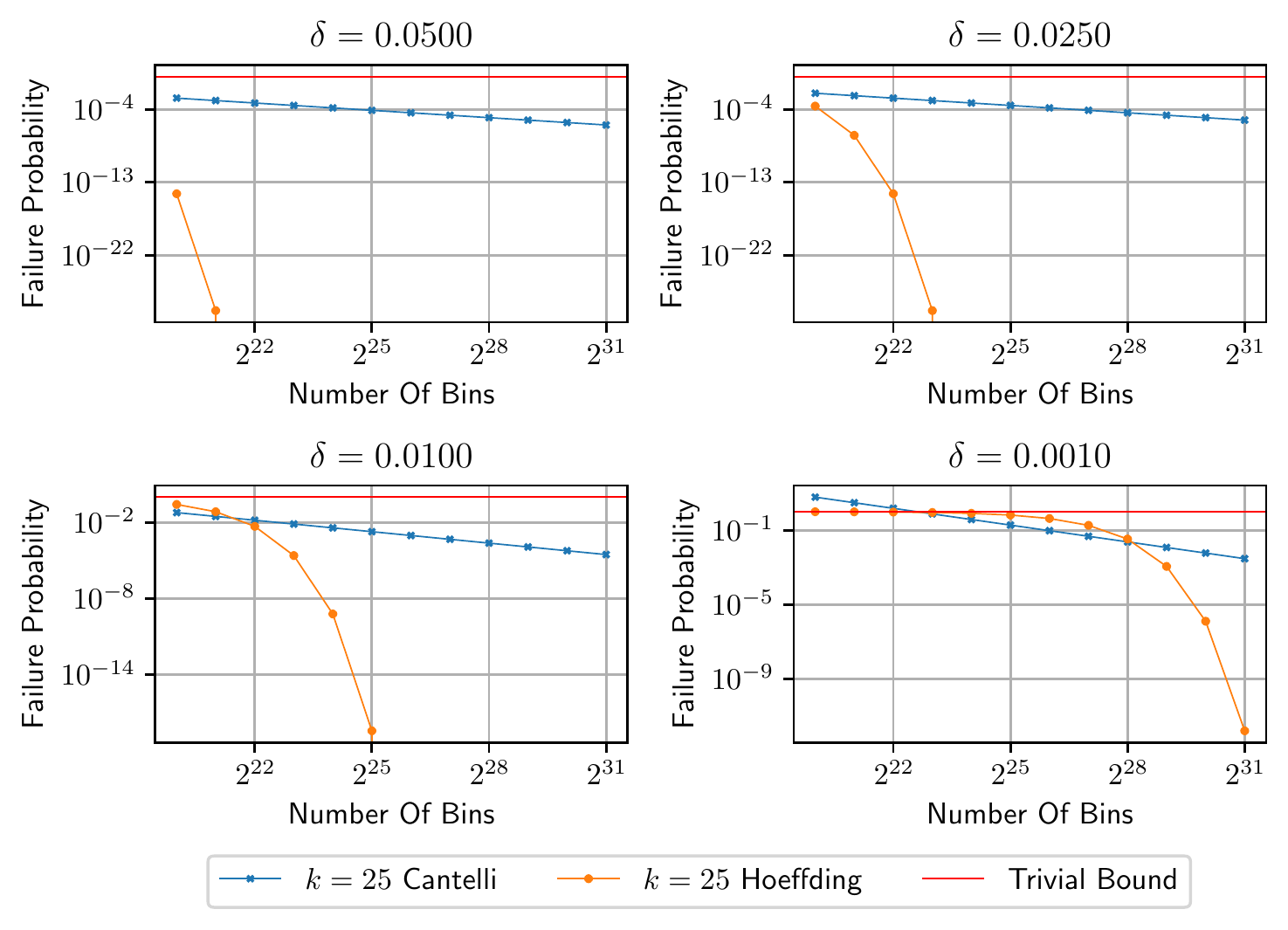}
    \vspace{-15pt}
    \caption{Comparison of the bounds.
        The $X$-axis denotes the number of bins ($m$).
        The logarithmic $Y$-axis denotes the probability that the \spare size, i.e $X$, is larger than $\Exp{X}(1+\delta)$.
        In each sub-figure, the choice of $\delta$ is specified in the title. When an achieved upper bound is greater than 1 we refer to it as \emph{trivial}.
    }\label{fig:concentration}
    % \vspace{-5pt}
\end{figure*}

While the above holds for all $n$, the Hoeffding bound in~\Cref{eq:deviation} yields an improved failure bound for large $n$, compared to the Cantelli bound part of~\Cref{eq:deviation}.
\Cref{fig:concentration} compares both bounds for different $\delta$ values as a function of the number of bins, $m=n/k$, for our prototype PD implementation's parameter choice of $k=25$.
For small values of $n$ (and hence $m$) and relatively small values of $\delta$, Hoeffding's bound achieve trivial bounds (bottom right of~\cref{fig:concentration}), while Cantelli's inequality yields non-trivial bounds.
Hoeffding's bound, however, is exponentially better for larger values of $n$ and $\delta$.
In particular, for $n\geq 2^{28}$ and  $\delta = \tfrac{1}{80}$, the Hoeffding bound yields a \PF failure probability smaller than $2^{-30}$ for our PD prototype's choice of $k=25$.

\subsection{Memory Access per Operation}\label{sec:prefix-ma}

\subsubsection{Queries}

We prove that every query requires a single memory access (i.e., reads one cache line) with probability at least $1-1/\sqrt{2\pi k}$, where $k$ denotes the capacity of a bin.
The proof deals separately with positive queries (i.e. $\query(x)$ for $x\in\DD$) and negative queries (i.e., $\query(x)$ for $x\notin\DD$).

\paragraph{Negative Queries}
\noindent
The following theorem deals with negative queries.
\begin{theorem}\label{thm:pf-no-query}
    A negative query in the \PF is forwarded to the \spare with probability at most
    \begin{align*}
        \Pr{\Bin{n}{p} = k+1}\leq \cgam\;.
    \end{align*}
\end{theorem}
\begin{proof}
    Consider a negative query with a key $x\notin\DD$.
    Let $i\triangleq \bin(x)$.
    Let $\fp_{(1)}\leq \fp_{(2)} \leq \cdots \leq \fp_{(B_i)}$
    denote the set of mini-fingerprints that are mapped to bin $i$ in
    non-decreasing order.
    Define two random variables $\varphi_i, \psi_i$:
    \begin{align*}
        \varphi_i\triangleq
        \begin{cases}
            \su-1     & \text{If } B_i\leq k \\
            \fp_{(k)} & \text{Otherwise.}
        \end{cases}
        \quad\psi_i\triangleq
        \begin{cases}
            1                 & \text{If } B_i\leq k \\
            {\ZZ}^{B_i}_{(k)} & \text{Otherwise.}
        \end{cases}
    \end{align*}
    where $\ZZ^{\ell}_{(k)}$ denotes the $k$'th smallest order statistic out of $\ell$
    i.i.d. random variables uniformly distributed in the interval $[0,1]$.

    The~\nameref{inv:prefix} implies that $\query(x)$ is forwarded to the \spare if and only if:
    \begin{enumerate*}
        \item $\bin(x)$ overflowed, and
        \item $\fp(x)$ is larger than the largest fingerprint currently stored in $\bin(x)$.
    \end{enumerate*}
    These two conditions are equivalent to the condition that $\fp(x)>\varphi_i$.
    Thus, it suffices to bound $\Pr{\fp(x)>\varphi_i}$.
    \begin{claim}\label{claim:uni exp}
        \begin{align*}
            \Pr{\fp(x)\leq\varphi_i} = \frac{\Exp{\varphi_i} +1}{s}
        \end{align*}
    \end{claim}
    \begin{proof}
        \renewcommand{\qedsymbol}{\ensuremath{\blacksquare}}
        \begin{align*}
            \Pr{\fp(x)\leq\varphi_i} & =
            \sum_{j=0}^{s-1}\sum_{h=j}^{s-1} \Pr{\fp(x) = j \wedge \varphi_i =h} \\&=            \sum_{j=0}^{s-1}\sum_{h=j}^{s-1} \frac{1}{s}\cdot \Pr{\varphi_i =h} \\&=
            \frac{1}{s}\sum_{j=0}^{s-1} (j+1)\cdot \Pr{\varphi_i =j} =
            \frac{\Exp{\varphi_i} +1}{s}
        \end{align*}
        Note that we relied on the uniform distribution of
        mini-fingerprints being independent and uniformly distributed
        in $[s]$ (independence implies that $\fp(x)$ and $\varphi_i$
        are independent).
    \end{proof}

    \begin{observation}\label{obs:phi}
        If $B_i>k$, then $\varphi_i$ and $\floor*{s\cdot \psi_i}$ are identically distributed.
    \end{observation}
    \begin{claim}\label{claim:phipsi}
        \begin{align*}
            \frac{\Exp{\varphi_i} +1}{s} \geq \Exp{\psi_i}
        \end{align*}
    \end{claim}
    \begin{proof}
        \renewcommand{\qedsymbol}{\ensuremath{\blacksquare}}
        We use the law of total expectation on the partition $\set{B_i\leq k}, \set{B_i> k}$, i.e.:
        \begin{align*}
            \Exp{\psi_i} =
            \Exp{\psi_i \mid B_i \leq k}\cdot \Pr{B_i \leq k} +
            \Exp{\psi_i \mid B_i > k}\cdot \Pr{B_i > k}
        \end{align*}
        Note that
        \begin{align*}
            \Exp{\psi_i \mid B_i \leq k} & = 1 =
            \Exp{\frac{\varphi_i +1}{s} \mid B_i \leq k} \\
            \Exp{\psi_i \mid B_i > k}    & \leq
            \Exp{\frac{\floor*{s\cdot \psi_i} +1}{s}\mid B_i > k} %
            =\Exp{\frac{\varphi_i+1}{s}\mid B_i > k}\;,
        \end{align*}
        and the claim follows.
    \end{proof}

    The random-variable $\ZZ^{\ell}_{(k)}$ (for $\ell\geq k$) is an order-statistic of uniform random variables, and satisfies the following property.
    \begin{proposition}{\cite[Section 4.6]{pitman1999probability}}
        For $\ell \geq k$,
        \begin{align*}
            \Exp{\ZZ^{\ell}_{(k)}} = \frac{k}{\ell+1}
        \end{align*}
    \end{proposition}

    \begin{claim}\label{claim:phi-expectation2}
        \begin{align*}
            \Exp{\psi_i} = \Pr{\Bin{n}{p} \leq k}+\frac{n}{(n+1)}\cdot \Pr{\Bin{n+1}{ p}> k+1}
        \end{align*}
    \end{claim}

    \begin{proof}%
        \renewcommand{\qedsymbol}{\ensuremath{\blacksquare}}
        \begin{align*}
            \Exp{\psi_i} & = \Exp{\psi_i \mid B_i \leq k}\cdot \Pr{B_i \leq k} + \Exp{\psi_i \mid B_i > k}\cdot \Pr{B_i > k} \\&=
            \Pr{B_i \leq k} + \underbrace{\Exp{\ZZ^{B_i}_{(k)} } \cdot \Pr{B_i> k}}_{S_1}
        \end{align*}
        Looking only at the second term ($S_1$):
        \begin{align*}
            S_1 & =
            \sum_{j>k}\Pr{B_i = j}\cdot \mathbb{E}[\ZZ^{j}_{(k)}]
            \\ &=
            \sum_{j>k}\Pr{B_i = j}\cdot \frac{k}{j+1}
            \\ &=
            \sum_{j>k}{n \choose j}p^{j}\cdot(1-p)^{n-j}\cdot \frac{k}{j+1}
            \\ &=
            \sum_{j>k}{n+1 \choose j+1}p^{j}\cdot(1-p)^{n-j}\cdot \frac{k}{n+1}
            \\ &=
            \frac{k}{p(n+1)}\cdot \sum_{j>k}{n+1 \choose j+1}p^{j+1}\cdot(1-p)^{n+1-(j+1)}
            \\&=\frac{n}{(n+1)}\cdot \Pr{\Bin{n+1}{ p}> k+1}\;,
        \end{align*}
        and the claim follows.
    \end{proof}

    \medskip\noindent
    The following claim allows us to shift from $\Bin{n+1}{ p}$ to $\Bin{n}{ p}$ using a ``rough'' bound.
    \begin{claim}\label{claim:identical1}
        \begin{align*}
            \frac{n}{(n+1)}\cdot \Pr{\Bin{n+1}{ p}> k+1} \geq \Pr{\Bin{n}{ p}> k+1}
        \end{align*}
    \end{claim}
    \begin{proof}
        \renewcommand{\qedsymbol}{\ensuremath{\blacksquare}}
        \begin{align*}
            \frac{n}{n+1}\cdot \Pr{\Bin{n+1}{p}>k+1} & =
            \frac{n}{n+1}  \sum_{j=k+2}^{\infty} \binom{n+1}{j}p^j\cdot\brak{1-p}^{n+1-k}
            \\
                                                     & =
            ( 1-p)\cdot \frac{n}{n+1} \sum_{j=k+2}^{\infty}  \binom{n+1}{j}p^j\cdot\brak{1-p}^{n-k}
            \\
                                                     & \overset{\star}{\geq}
            \sum_{j=k+2}^{\infty} \binom{n}{j}p^j\cdot\brak{1-p}^{n-k}
            \\
                                                     & =
            \Pr{\Bin{n}{p}>k+1}\;.
        \end{align*}
        We need to show that the ($\star$) inequality is true.
        For that, it suffices to show that
        \begin{align}
            (1-p)\cdot\frac{n}{n+1}\cdot \binom{n+1}{j} \geq \binom{n}{j}\;.\label{eq:npone5}
        \end{align}

        We prove~\Cref{eq:npone5} by showing its left-hand side divided by its right-hand side is larger than 1.
        \begin{align*}
            (1-p)\cdot\frac{n}{n+1}\cdot \frac{\binom{n+1}{j}}{\binom{n}{j}} =
            \frac{n-k}{n}\cdot\frac{n}{n+1}\cdot \frac{n+1}{n+1-j} =
            \frac{n-k}{n+1-j} \overset{n\geq j>k+1}{>} 1
        \end{align*}

    \end{proof}
    \begin{conclusion}\label{conclusion:z5}
        \begin{align*}
            \Exp{\psi_i} \geq 1-\Pr{\Bin{n}{ p}= k+1}
        \end{align*}
    \end{conclusion}
    \begin{proof}
        \renewcommand{\qedsymbol}{\ensuremath{\blacksquare}}
        \begin{align*}
            \Exp{\psi_i} & = \Pr{\Bin{n}{p} \leq k}+\frac{n}{(n+1)}\cdot \Pr{\Bin{n+1}{ p}> k+1} \\
                         & \geq
            \Pr{\Bin{n}{p} \leq k}+\Pr{\Bin{n}{ p}> k+1}                                         \\
                         & =1-\Pr{\Bin{n}{p}=k+1} \;.
        \end{align*}
    \end{proof}
    \medskip
    \noindent
    To complete the proof of \cref{thm:pf-no-query} we combine the above inequalities to obtain
    \begin{align*}
        \Pr{\fp(x)\leq\varphi_i}  \overset{\cref{claim:uni exp}}{=} \frac{\Exp{\varphi_i + 1}}{s} \overset{\cref{claim:phipsi}}{\geq}
        \Exp{\psi_i}              \overset{\Cref{conclusion:z5}}{\geq} 1-\Pr{\Bin{n}{ p}= k+1}
    \end{align*}
    The first part of the proof follows by taking the complement event.
    \begin{align*}
        \Pr{\fp(x)>\varphi_i} \leq \Pr{\Bin{n}{p} = k+1}
    \end{align*}
    Finally, we note that:
    \begin{align*}
        \Pr{\Bin{n}{p} = k+1} \leq \Pr{\Bin{n}{p} = k} \leq \cgam\;
    \end{align*}
    Where the last inequality is true by \cref{prop:mu}, and the theorem follows.
\end{proof}

\paragraph{Positive Queries}
We prove the following theorem.
\begin{theorem}[Spare access by positive queries]\label{thm:pos-query}
    A positive query in the \PF is forwarded to the \spare with probability at most $\cgam$.
\end{theorem}

\begin{proof}
    We begin with a short discussion that explains why the analysis of the number of memory accesses per negative query does not apply to the case of positive queries.
    Consider a "snapshot" of the \PF at time $t$ when the positive query $\query(y)$ is issued. Let $\sigma$ denote the sequence of $t$ operations (note that the last operation in $\sigma$ is $\query(y)$), and let $\DD_t$ denote the dataset consisting of the elements inserted in $\sigma$.
    Let $\FP$ denote the hash function, let $\bin(y)=i$.
    Our goal is to bound the probability that $\query(y)$ is forwarded to the \spare.
    Note that the probability that $\query(y)$ is forwarded to the \spare is nondecreasing in $\size{\DD_t}$. Hence, for the purpose of proving~\Cref{thm:pos-query}, we may assume that $\size{\DD_t}=n$.

    The analysis for negative queries does not hold for positive queries for the following reason.
    Indeed,
    \begin{align*}
        \Pr{\query(y)\text{ is forwarded to the }\spare} = \Pr{\fp(y) > \varphi_i}\;. %
    \end{align*}
    However, the random variables $\fp(y)$ and $\varphi_i$ are no longer independent (as $y$ may be one of the elements whose mini-fingerprints appears in the prefix).
    Hence, we can no longer argue that
    $\Pr{\fp(y) > \varphi_i}=\Exp{\varphi_i}$.

    \newcommand{\cv}{\hat{\chi}_i}
    \newcommand{\hv}{\hat{\varphi}_i}
    \newcommand{\hp}{\hat{\psi}_i}
    \newcommand{\up}{{\upsilon}_i}
    \newcommand{\OS}[1]{\mathtt{OrderStat}_{k}^{s}(#1)}

    We resolve this obstacle by "removing" $y$ as follows.
    We use the notation $\lset{\cdot}$ to denote a set with repetitions i.e. a \emph{multiset}. The multiplicity of $x$ in a multiset $A$ is the number of times $x$ appears in $A$. Note that every set is also a multiset in which every element has multiplicity $0$ or $1$.
    \begin{align*}
        A_y & \triangleq\set{x\in \DD_t\mid \bin(x)=\bin(y)}, & A^\prime_y & \triangleq A_y - \set{y}                       \\
        F_y & \triangleq\lset{\fp(x)\mid x\in A_y},           & F^\prime_y & \triangleq\lset{\fp(x)\mid x\in A^\prime_y}\;.
    \end{align*}
    Note that elements in $A_y$ are distinct (i.e., multiplicity is zero or one).
    However, $F_y$ is a multiset due to collisions introduced by different elements having identical fingerprints.
    \begin{definition}[Strict rank]
        Define the strict rank of $x$ with respect to a multiset $A$ as follows.
        \begin{align*}
            \rank(x, A)\triangleq\size{\lset{z\in A\mid z<x}}\;.
        \end{align*}
        Note that $x$ does not have to belong to $A$, and that $\rank(x,A)=0$ means that $x\leq \min_{a\in A} a$. Moreover, the strict rank counts the sum of the multiplicities of elements in $A$ that are strictly smaller than $x$.
    \end{definition}
    Let $A+\lset{x}$ denote the multiset obtained from $A$ by increasing the multiplicity of $x$ by one. We refer, informally, to $A+\lset{x}$ as \emph{adding $x$ to $A$}.

    The following observation states that adding $x$ to $A$ does not change the strict rank of $x$ with respect to $A$.
    \begin{observation}\label{obser:rank}
        For any multiset $A$ and an element $x$ we have
        \begin{align*}
            \rank(x, A) = \rank(x, A+\lset{x})
        \end{align*}
    \end{observation}
    \noindent
    We also define
    \begin{align*}
        \cv\triangleq
        \begin{cases}
            s-1                 & \text{If } B_i\leq k \\
            \min_k {F^\prime_y} & \text{Otherwise.}
        \end{cases} &  &
        \hp\triangleq
        \begin{cases}
            1                     & \text{If } B_i\leq k \\
            {\ZZ}^{B_i - 1}_{(k)} & \text{Otherwise.}
        \end{cases}
    \end{align*}
    Where for a multiset $S$, we define $\min_j(S)$ to be the $j$'th smallest element in this multiset. Note that $\min_j(S)$ is not necessarily the element of strict rank $j-1$. Indeed, consider the multiset $S$ consisting of $n\geq j$ repetitions of the same element $x$, then $x=\min_j (S)$ but $\rank(x,S)=0$.
    The~\nameref{inv:prefix} says that $\min_1(A_y),\ldots, \min_k(A_y)$ are stored in $\bin(y)$ if the bin is full.

    \begin{claim}\label{claim:reduce to chi}
        \begin{align*}
            \Pr{\query(y)\text{ is forwarded to the }\spare}  = \Pr{\fp(y)> \cv}
        \end{align*}
    \end{claim}
    \begin{proof} %
        \renewcommand{\qedsymbol}{\ensuremath{\blacksquare}}
        We already noted that
        \begin{align*}
            \Pr{\query(y)\text{ is forwarded to the }\spare} = \Pr{\fp(y) > \varphi_i}\;.
        \end{align*}
        We are left with showing that
        \begin{align*}
            \Pr{\fp(y) > \varphi_i} = \Pr{\fp(y)> \cv} %
        \end{align*}
        Note that
        \begin{align*}
            \rank(\fp(y),F_y) =\rank(\fp(y),F^\prime_y)\;,
        \end{align*}
        which is a special case of~\Cref{obser:rank}. Therefore,
        \begin{align*}
            \set{\fp(y)> \varphi_i}=
            \set{\rank(\fp(y),F_y) \geq k} =
            \set{\rank(\fp(y),F^\prime_y) \geq k}=
            \set{\fp(y)> \cv}\;.
        \end{align*}
        Therefore, the two events $\set{\fp(y)> \varphi_i}, \set{\fp(y)> \cv}$ are equal, which completes the proof.
    \end{proof}
    \begin{claim}\label{claim:reduce to psi}
        \begin{align*}
            \Pr{\fp(y)\leq \cv} \geq \Exp{\psi_i}
        \end{align*}
    \end{claim}

    \begin{proof} %
        \renewcommand{\qedsymbol}{\ensuremath{\blacksquare}}
        \begin{align*}
            \Pr{\fp(y)\leq \cv} \overset{\cref{claim:uni exp}}{=} \frac{\Exp{\cv} + 1}{s}\overset{\cref{claim:phipsi}}{\geq} \Exp{\hp}\overset{\star}{\geq} \Exp{\psi_i}
        \end{align*}
        We prove the last inequality by case analysis on $B_i$.
        \begin{align*}
             & \Exp{\hp\mid B_i\leq k} = 1 =\Exp{\psi_i\mid B_i\leq k}                \\
             & \Exp{\hp\mid B_i> k} = \Exp{{\ZZ}^{B_i - 1}_{(k)}}=\frac{k}{B_i - 1} >
            \frac{k}{B_i} = \Exp{{\ZZ}^{B_i}_{(k)}} =\Exp{\psi_i\mid B_i > k}
        \end{align*}
    \end{proof}

    We now complete the proof of~\cref{thm:pos-query}.
    By \cref{claim:reduce to chi,claim:reduce to psi} we get
    \begin{align*}
        \Pr{\query(y)\text{ is {\bf{not}} forwarded to the }\spare}\geq \Exp{\psi_i}\;.
    \end{align*}
    By~\Cref{conclusion:z5},
    we get that
    \begin{align*}
        \Exp{\psi_i}\geq 1-\Pr{\Bin{n}{ p}= k+1} \geq 1-\cgam
    \end{align*}
    and therefore conclude that
    \begin{align*}
        \Pr{\query(y)\text{ is {\bf{not}} forwarded to the }\spare}\geq 1-\cgam
    \end{align*}
    which completes the proof.
\end{proof}

\subsubsection{Insertions}
The number of insert operations that require more than one memory access equals the number of elements that are stored in the
\spare. By \Cref{claim:fail},  with probability $\cProbVar$, all but at most $\frac{1.1\cdot n}{\cdom}$ insertions require one memory access.

\subsection{False Positive Rate} \label{sec:filter-fpp}
In this section, we analyze the false positive rate of the \PF.
Our analysis assumes that the \spare is constructed with false positive rate $\eps'$ for a set of size at most $1.1 \cdot \expectation{}{X}$ and that failure did not occur (i.e. $X\leq 1.1 \cdot \expectation{}{X}$).

A false positive event for $\query(y)$ (for $y\in \UU\setminus \DD)$ is contained in the union of two events:
\begin{enumerate*}[label=(\Alph*)]
    \item
    Fingerprint collision, namely, there exists $x\in\DD$ such that $\FP(y)=\FP(x)$.
    \item
    A fingerprint collision did not occur and the \spare is accessed to process $\query(y)$ and answers "yes".
\end{enumerate*}
Let $\eps_1\triangleq \Pr{A}$ and $\delta_2\triangleq \Pr{B}$.
To bound the false positive rate it suffices to bound $\eps_1+\delta_2$.

\begin{claim} \label{claim:eps1}
    $\eps_1 \leq \frac{n}{m\cdot s}$ and $\delta_2\leq \frac{\eps'}{\sqrt{2\pi k}}$.
\end{claim}
\begin{proof}
    \renewcommand{\qedsymbol}{\ensuremath{\blacksquare}}
    Fix $y \in \UU \setminus \DD$.	The probability that $\FP(x)=\FP(y)$ is $\frac{1}{m\cdot s}$ because $\FP$ is chosen from a family of $2$-universal hash functions whose range is $[m\cdot s]$.
    The bound on $\eps_1$ follows by applying a union bound over all $x\in\DD$.
    To prove the bound on $\delta_2$, note that every negative query to the \spare generates a false positive with probability at most $\eps'$.
    However, not every query is forwarded to the \spare.
    In~\cref{thm:pf-no-query} we bound the probability that a query is forwarded to the spare by $1/\sqrt{2 \pi k}$, and the claim follows.
\end{proof}

\begin{corollary}
    The false positive rate of the \PF is at most $\frac{n}{m \cdot s}+ \frac{\eps'}{\sqrt{2\pi k}}$.
\end{corollary}

\section{Evaluation}\label{sec:evaluation}

In this section, we empirically compare the \PF to other
state-of-the-art filters with respect to several metrics:
space usage and false positive rate (\cref{sec:eval:space}),
throughput of filter operations at different loads (\cref{sec:eval:filling}), and build time, i.e., overall time to insert $n$ keys into an empty filter (\cref{sec:build time}).

\subsection{Experimental Setup}

\paragraph{Platform.} We use an Intel Ice Lake CPU (Xeon Gold 6336Y CPU @ 2.40\,GHz), which has per-core, private 48\,KiB L1 data caches and 1.25\,MiB L2 caches, and a shared 36\,MiB L3 cache.
The machine has 64\,GiB of RAM. Code is compiled using GCC 10.3.0 and run on Ubuntu 20.04.
Reported numbers are medians of 9 runs.
Medians are within $-3.41$\% to $+4.4\%$ of the corresponding averages, and 98\% of the measurements are less than 1\% away from the median.

\paragraph{False positive rate ($\eps$).}
Our \PF prototype supports a false positive rate of $\eps=0.37\% \lessapprox 2^{-8}$ (\cref{sec:pd-params}). However, not all filters support this false positive rate (at least not with competitive speed), thus making an apples-to-apples comparison difficult. We therefore configure each filter with a false positive rate that is as close as possible to $2^{-8}$ without deteriorating its speed. \Cref{sec:eval:filters} expands on these false positive rate choices.

\paragraph{Dataset size ($n$).}
We use a dataset size (maximum number of keys that can be inserted into the filter) of $n = 0.94 \cdot 2^{28}$ (252\,M), which ensures that filter size exceeds the CPU's cache capacity. We purposefully do not choose $n$ to be a power of $2$ as that would unfairly disadvantage some
implementations, for the following reason.

Certain hash table of fingerprints designs, such as the cuckoo and vector quotient filter, become ``full'' and start failing insertions (or slow down by orders of magnitude) when the load factor of the underlying hash table becomes ``too high'' (\cref{table:the-space}). For instance, the cuckoo filter occasionally fails if its load factor exceeds 94\%~\cite{graf2020xor}. It should thus size its hash table to fit $n/0.94$ keys. Unfortunately, the fastest implementation of the cuckoo filter (by its authors~\cite{dblp:conf/conext/fanakm14}) cannot do this. This implementation is what we call \emph{non-flexible}: it requires $m$, the number of hash table bins, to be a power of $2$, so that the code can truncate a value to a bin index with a bitwise-ANDs instead of an expensive modulo operation. The implementation uses bins with capacity 4 and so its default $m$ is $n/4$ rounded up to a power of 2. The maximal load factor when $n$ is a power of 2 is thus 1, so it must double $m$ to avoid failing, which disadvantages it in terms of space usage.

Our choice of an $n$ close to a power of 2 avoids this disadvantage. When created with our $n = 0.94 \cdot 2^{28}$, the cuckoo filter uses $m=2^{28}/4$ bins, and thus its load factor in our experiments  never exceeds the supported load factor of 94\%.

\paragraph{Implementation issues.}
We pre-generate the sequences of operations (and arguments) so that measured execution times reflect only filter performance.
While the keys passed to filter operations are random, we do not remove any internal hashing in the evaluated filters, so that our measurements reflect realistic filter use where arguments are not assumed to be uniformly random.
All filters use the same hash function, by Dietzfelbinger~\cite[Theorem 1]{Dietzfelbinger2018}.

\subsubsection{Compared Filters} \label{sec:eval:filters}
We evaluate the following filters:

\paragraph{Bloom filter (BF-$x[k=\cdot]$).}
We use an optimized Bloom filter implementation~\cite{graf2020xor}, which uses two hash functions to construct $k$ hash outputs.
The parameter $x$ denotes the number of bits per key.%

\paragraph{Cuckoo Filter (CF-$x$).} ``CF-$x$'' refers to a \CF
with fingerprints of $x$ bits with buckets of $4$ fingerprints (the bucket size recommended by the \CF authors).
The \CF's false positive rate is dictated by the fingerprint length, but competitive speed requires fingerprint length that is a multiple of 4.
Thus, while a CF-11 has a false positive rate of $\approx 2^{-8}$, it is very slow due non-word-aligned bins or wastes space to pad bins.
We therefore evaluate CF-8 and CF-12, whose false positive rates of 2.9\% and 0.18\%, respectively, ``sandwich'' the \PF's rate of $0.37\%$.

We evaluate two \CF implementations: the authors' implementation~\cite{CFimpl}, which is non-flexible, and a flexible implementation~\cite{FastFilterImpl}, denoted by a
``\textsf{-Flex}'' suffix, which does not require the number of bins in the hash table to be a power of 2.

\paragraph{Blocked Bloom filter (BBF).} We evaluate two implementations  of  this filter: a non-flexible implementation (i.e., the bit vector length is a power of $2$) from the \CF
repository and a
flexible version, \BBFx, taken from~\cite{graf2020xor}.
The false positive rate of these implementations cannot be tuned, as they set a fixed number of bits on insertion, while in a standard \BF the number of bits set in each insertion depends on the desired false positive rate.
It is possible to control the false positive rate by decreasing the load (i.e., initializing the \BBF with a larger $n$ than our experiments actually insert), but then its space consumption would be wasteful.
We therefore report the performance of the \BBF implementations with the specific parameters they are optimized for.

\paragraph{TwoChoicer (TC).} This is our implementation of the \VQF~\cite{pandey2021vector}.
Its hash table bins are implemented with our pocket dictionary, which is faster than the bin implementation of the \VQF authors~\cite{VQFImpl}.%
\footnote{In our throughput evaluation~(\cref{sec:eval:filling}),
    the throughput of the TC is higher than the \VQF by at least $1.26\times$ in negative queries at any load, and is comparable at insertions and positive queries.}
We use the same bin parameters as in the original \VQF implementation:
a 64-byte PD with a capacity of $48$ with parameters $Q = 80$ and $R = 8$.
Setting $R=8$ leads to an empirical false positive rate of $0.44$\%.

\paragraph{Prefix-Filter (PF[Spare]).}
We use a \PF whose bin table contains $m = \frac{n}{0.95\cdot k}$ bins (see~\cref{sec:the-pf:which-spare}).
We evaluate the \PF with three different implementations of the \spare: a \BBFx, CF-12-Flex, and TwoChoicer, denoted PF[BBF-Flex], P[CF12-Flex], and PF[TC], respectively.
Let $n'$ denote the \spare's desired dataset size derived from our analysis~(\cref{sec:the-pf:spare-params}).
We use a \spare dataset size of $2n', n'/0.94, n'/0.935$ for
PF[BBF-Flex], PF[CF12-Flex], and PF[TC], respectively.
The purpose of the BBF-Flex setting is to obtain the desired false positive rate, and the only purpose of the other settings is to avoid failure.

\paragraph{Omitted filters.}
We evaluate but omit results of the Morton~\cite{breslow2018morton} and
quotient filter~\cite{bender2012thrash}, because they are strictly worse than the vector quotient filter (TC).%
\footnote{In addition, the Morton filter is known to be similar or worse than the \CF on Intel CPUs such as ours~\cite{breslow2018morton}.}
This result is consistent with prior work~\cite{pandey2021vector}.

\subsection{Space and False Positive Rate} \label{sec:eval:space}

We evaluate the size of the filters by inserting $n$ random keys and measuring the filter's space consumption, reporting it in bits per key.
We then measure the false positive rate by performing $n$ random queries (which due to the universe size are in fact negative queries) and measuring the fraction of (false) positive responses.
\Cref{table:fpp-server} shows the results.
We also compare the space use (bits/key) of each filter to the information theoretic minimum for the same false positive rate, examining both additive and multiplicative differences from the optimum.

Comparing space use of filters with different false positive rates is not meaningful, so we compare each filter's space overhead over the information theoretic minimum for the same false positive rate.
Except for the Bloom and blocked Bloom filters, which have multiplicative overhead, the relevant metric is the additive difference from the optimum.
We see that in practice, all filters use essentially the same space, $3.4$--$4$ bits per key more than the optimum.

The \PF's space use and false positive rates are nearly identical regardless of whether BBF-Flex, CF-12-Flex, or TC is used as a \spare.
This holds despite the fact that as standalone filters, the false positive rates and space use of these filters differs considerably.
This result empirically supports our analysis that the impact of the \spare on the \PF's overall space use and false positive rate is negligible.

\begin{table}
    \scalebox{0.94}{
        \begin{tabular}{l|lllll}
            \toprule
            \textbf{Filter} & \thead{\textbf{Error}                                \\ \textbf{(\%)}} & \textbf{Bits/key} & \thead{\textbf{Optimal} \\\textbf{bits/key}} & \textbf{Diff.} & \textbf{Ratio} \\
            \midrule
            CF-8            & 2.9163                & 8.51  & 5.10  & 3.41 & 1.669 \\
            CF-8-Flex       & 2.9175                & 8.51  & 5.10  & 3.41 & 1.669 \\
            CF-12           & 0.1833                & 12.77 & 9.09  & 3.67 & 1.404 \\
            CF-12-Flex      & 0.1833                & 12.77 & 9.09  & 3.67 & 1.404 \\
            CF-16           & 0.0114                & 17.02 & 13.10 & 3.92 & 1.299 \\
            CF-16-Flex      & 0.0114                & 17.02 & 13.10 & 3.92 & 1.299 \\
            \midrule
            PF[BBF-Flex]    & 0.3723                & 12.13 & 8.07  & 4.06 & 1.503 \\
            PF[CF-12-Flex]  & 0.3797                & 11.64 & 8.04  & 3.60 & 1.447 \\
            PF[TC]          & 0.3917                & 11.55 & 8.00  & 3.56 & 1.445 \\
            \midrule
            BBF             & 2.5650                & 8.51  & 5.28  & 3.23 & 1.610 \\
            BBF-Flex        & 0.9424                & 10.67 & 6.73  & 3.94 & 1.585 \\
            BF-8[k=6]       & 2.1611                & 8.00  & 5.53  & 2.47 & 1.446 \\
            BF-12[k=8]      & 0.3166                & 12.00 & 8.30  & 3.70 & 1.445 \\
            BF-16[k=11]     & 0.0460                & 16.00 & 11.09 & 4.91 & 1.443 \\
            \midrule
            TC              & 0.4447                & 11.41 & 7.81  & 3.60 & 1.460 \\
            \bottomrule
        \end{tabular}
    }
    \caption{%
    Filter false positive rate and space use: empirically obtained results (for $n=2^{28}\cdot 0.94$) and comparison to the information theoretic minimum.
    }
    \label{table:fpp-server}
\end{table}

\begin{figure*}[t]
    \centering
    \includegraphics[width=\textwidth]{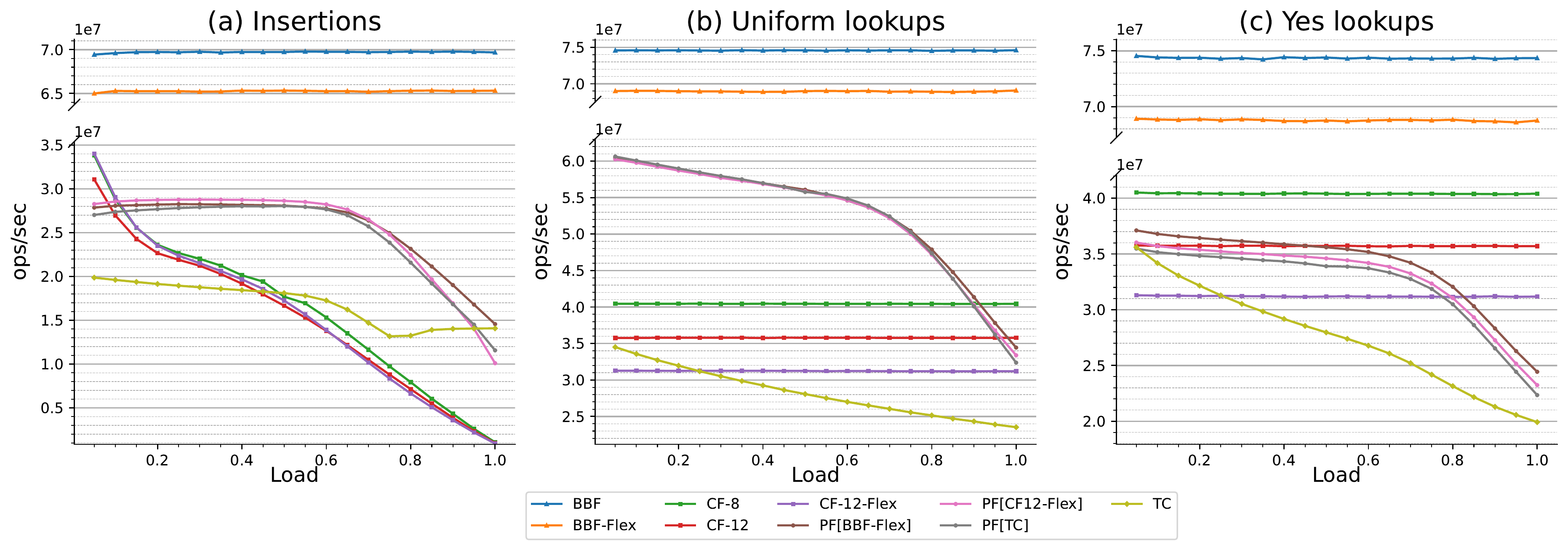}
    \vspace{-20pt}
    \caption{Throughput comparison as filter load increases ($X$-axis shows the fraction of $n=2^{28}\cdot 0.94$ elements
    inserted so far).
    }\label{fig-perf}
\end{figure*}

\subsection{Throughput} \label{sec:eval:filling}

We evaluate the throughput of filter operations as the load (fraction of keys inserted out of $n$) gradually increases in increments of 5\%. We report the throughput of insertions, negative queries, and positive queries.
Prior work~\cite{pandey2021vector,bender2012thrash,breslow2018morton,dblp:conf/conext/fanakm14} uses similar methodology to evaluate the impact of filter load balancing.

Each experiment consists of 20 rounds.
Each round consists of three operation sequences whose throughput is measured.
The round begins with a sequence of $0.05n$ insertions of uniformly random 64-bit keys.
We then perform a sequence of $0.05n$ queries of uniformly random keys (which due to the $2^{64}$ universe size are, in fact, negative queries).
The round ends with a sequence of $0.05n$ positive queries by querying for a randomly permuted sample of $0.05n$ keys that were inserted in some previous round.
Key generation does not affect throughput results, as we only time round executions, whereas the insertion sequence is pre-generated and the positive query sequences are generated between rounds.

\Cref{fig-perf} shows the throughput (operations/second) of
insertions, uniform (negative) queries, and positive queries for each load (i.e., round).

We omit the \BF from the plots, as it is the slowest in queries, and only faster than \CF in high-load insertions.
The \BBF outperforms all other filters by about $2\times$ for all operations and all loads.
But the \BBF has significantly worse space efficiency (which due to implementation limitations is reflected in $3\times$--$6\times$ higher false positive rates, see~\cref{sec:eval:filters}).
We therefore do not consider it further.

\paragraph{Negative queries.}
Compared to CF-12-Flex and TC, whose false positive rate is similar to the \PF's, the \PF has higher negative query throughput at all loads.
In fact, the \PF has higher negative query throughput than the CF-12 for all loads up to 95\%: the \PF's throughput is higher by

55\%, 40\% and 2.8\% for loads of 50\%, 70\%, and 90\%,
respectively (same result for all \spare implementation up to $\pm 3\%$).
The main reason is that CF and TC negative queries incur two cache misses, whereas \PF negative queries that do not access the \spare incur only a single cache miss.
Both \PF and TC, whose bins are implemented as pocket dictionaries, exhibit a gradual decline in negative query throughput as load increases.
This decline occurs because as bins become fuller, the query ``cutoff'' optimization becomes less effective.
We also see the space vs.
speed trade-off in a non-flexible implementation.
The query throughput of CF-12 and BBF is higher than that of their flexible counterparts by 14.5\% and 8.1\%, respectively.
In exchange for this, however, the non-flexible implementation may use $2\times$ more space than a flexible one.

\paragraph{Insertions.}
The \PF has about $1.25\times$ higher insertion throughput than TC for loads up to roughly 85\%.
Subsequently, the \PF's insertion throughput gradually decreases, but it becomes lower than TC's only at 95\% load.
TC's throughput degrades when the load exceeds 50\% due to its insertion shortcut optimization, because at these loads a larger number of TC insertions require access to two bins.

The \PF has higher insertion throughput than CF-12 and CF-8 for all loads above 10\%. The reason is that the \CF's insertion throughput declines by orders of magnitude as load increases, whereas the \PF's throughput is stable up to 60\% load and subsequently declines by only $1.9\times$--$2.8\times$, depending on the \spare implementation.

\paragraph{Positive queries.}
At 100\% load, the \PF's positive query throughput is lower by $28\%-40\%$ compared to CF-12-Flex and by $46\%-60\%$ compared to CF-12.
The \PF is consistently faster than TC, e.g., by 12\%-22\% (depending on the \spare implementation) at 100\% load.

\subsection{Build Time}\label{sec:build time}

In this experiment, we measure filter \emph{build time}: the time it takes to insert $n$ random keys into an initially empty filter. This is a very common workload for filters, e.g., whenever the filter represents an immutable dataset~\cite{KVstoreAnalysis,RocksDBAnalysis,conway_et_al:LIPIcs:2018:9043,SplinterDB,dayan2017monkey,PebblesDB,rocksdb,LSM-SSD,Chucky,Dostoevsky,chang2008bigtable,Cassandra,bLSM,cLSM}.

\begin{figure}[t]
    \centering
    \includegraphics[width=0.9\columnwidth]{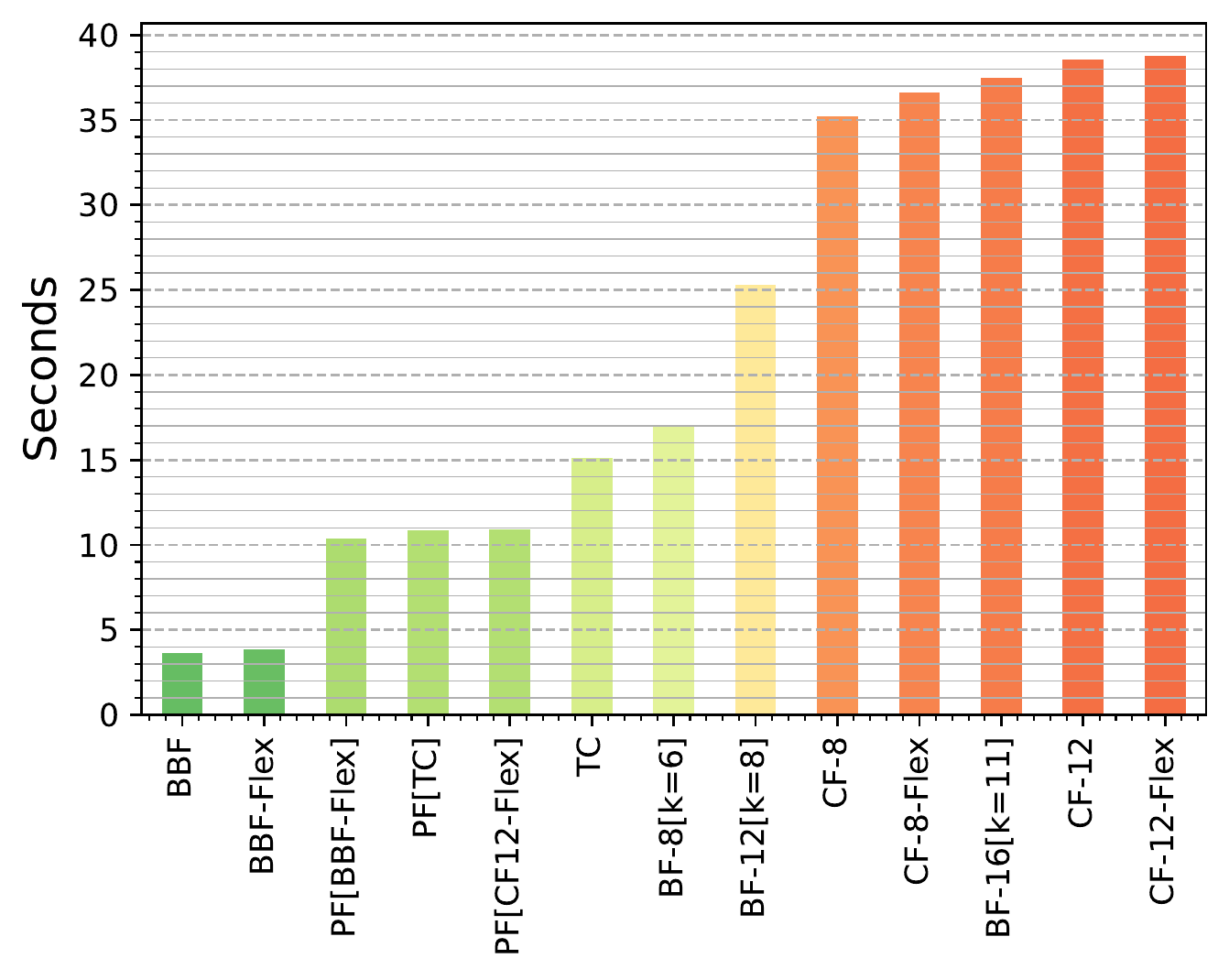}
    \vspace{-10pt}
    \caption{Filter build time (seconds) for
        $n=0.94 \cdot 2^{28}$.
    }
    \label{fig:build-time}
\end{figure}

\Cref{fig:build-time} shows the build time of the evaluated filters.
Build time reflects average of insertion speed as load increases.
Therefore, the \PF's high insertion throughput at loads below $80\%$, which decreases only gradually at higher loads, translates into overall faster build time than all CF configurations and TC---by $>3.2\times$ and $1.39\times$--$1.46\times$, respectively.
The choice of the \spare implementation has only a minor influence on the \PF's build time (a difference of $5.6$\% between the \PF's worst and best build times).

The \CF's precipitous drop in insertion throughput as load increases translates into build times similar to BF-16[$k=11$], whose operations perform 11 memory accesses.
In other words, with respect to build time, the \CF behaves similarly to a filter whose insertions cost about 11 cache misses.

\section{Conclusion}

We propose the \PF, an incremental filter that offers (1) space efficiency comparable to state-of-the-art dynamic filters; (2) fast queries, comparable to those of the \CF; and (3) fast insertions, with overall build times faster than those of the \VQF and \CF. Our rigorous analysis of the \PF's false positive rate and other properties holds both for practical parameter values and asymptotically. We also empirically evaluate the \PF and demonstrate its qualities in practice.

\begin{acks}
    This research was funded in part by the Israel Science Foundation
    (grant 2005/17) and the Blavatnik Family Foundation.
\end{acks}

% \balance
\bibliographystyle{ACM-Reference-Format}
\bibliography{filters}

\end{document}